\documentclass[a4paper,11pt]{article}
\pdfoutput=1 

\usepackage{jinstpub} 
\usepackage{subcaption}
\usepackage{upgreek}
\usepackage{doi}

\title{Charge collection properties in an irradiated pixel sensor built in a thick-film HV-SOI process}

\author[a,1]{B. Hiti\note{Corresponding author.}}
\author[a]{\hspace{-0.15cm}, V. Cindro}
\author[a]{\hspace{-0.15cm}, A. Gori\v{s}ek}
\author[c]{\hspace{-0.15cm}, T. Hemperek}
\author[c,2]{\hspace{-0.15cm}, T. Kishishita\note{Now Institute of Particle and Nuclear Science, KEK High Energy Accelerator Research Organization.}}
\author[a]{\hspace{-0.15cm}, G. Kramberger}
\author[c]{\hspace{-0.15cm}, \\H. Kr\"uger}
\author[a]{\hspace{-0.15cm}, I. Mandi\'{c}}
\author[a,b]{\hspace{-0.15cm}, M. Miku\v{z}}
\author[c]{\hspace{-0.15cm}, N. Wermes}
\author[a]{\hspace{-0.15cm}, M. Zavrtanik}

\affiliation[a]{Jo\v{z}ef Stefan Institute,\\Jamova 39, Ljubljana, Slovenia}
\affiliation[b]{University of Ljubljana, Faculty of Mathematics and Physics,\\Jadranska 19, Ljubljana, Slovenia}
\affiliation[c]{University of Bonn, Physikalisches Institut,\\Nu{\ss}allee 12, Bonn, Germany}

\emailAdd{bojan.hiti@ijs.si}
\arxivnumber{1701.06324}

\abstract{
Investigation of HV-CMOS sensors for use as a tracking detector in the ATLAS experiment at the upgraded LHC (HL-LHC) has recently been an active field of research. A potential candidate for a pixel detector built in Silicon-On-Insulator (SOI) technology has already been characterized in terms of radiation hardness to TID (Total Ionizing Dose) and charge collection after a moderate neutron irradiation.
In this article we present results of an extensive irradiation hardness study with neutrons up to a fluence of $1\times 10^{16}\,\mathrm{n}_{\mathrm{eq}}/\mathrm{cm}^2$. Charge collection in a passive pixelated structure was measured by Edge Transient Current Technique (E-TCT). The evolution of the effective space charge concentration was found to be compliant with the acceptor removal model, with the minimum of the space charge concentration being reached after $5\times 10^{14}\,\mathrm{n}_{\mathrm{eq}}/\mathrm{cm}^2$. 
An investigation of the in-pixel uniformity of the detector response revealed parasitic charge collection by the epitaxial silicon layer characteristic for the SOI design. The results were backed by a numerical simulation of charge collection in an equivalent detector layout.
}


\begin{document}
\maketitle
\flushbottom

\section{Introduction}
\label{sec:intro}

Intensive investigations of the possibility to produce particle tracking detectors for experiments at upgraded LHC (HL-LHC) \cite{ATLAS-LOI} using technology for commercial integrated CMOS circuits \cite{Dierickx} are currently ongoing at several research institutions throughout the world. 
The potential of the CMOS technology offers production of fully monolithic particle detectors \cite{Kemmer}, which would enable a smaller pixel size, simpler assembly without costly interconnections between sensors and readout electronics, and consequently less material in the tracking volume \cite{ATLAS-pixel, Kastli, Aamodt, Wermes-book}.
Manufacturing detectors in commercial fabrication plants should also result in faster production, larger number of vendors and consequently lower cost.

\begin{figure}%
\centering
\begin{subfigure}[b]{0.34\textwidth}
\includegraphics[width=\columnwidth]{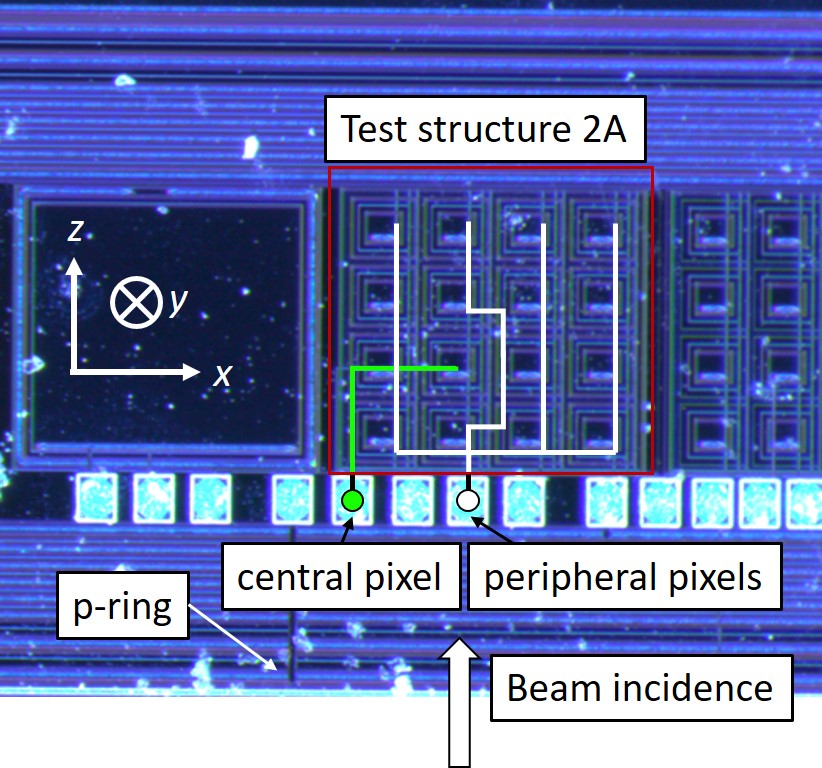}%
\caption{}
\label{fig:photo-array}
\end{subfigure}
\begin{subfigure}[b]{0.6\textwidth}
\includegraphics[width=\columnwidth]{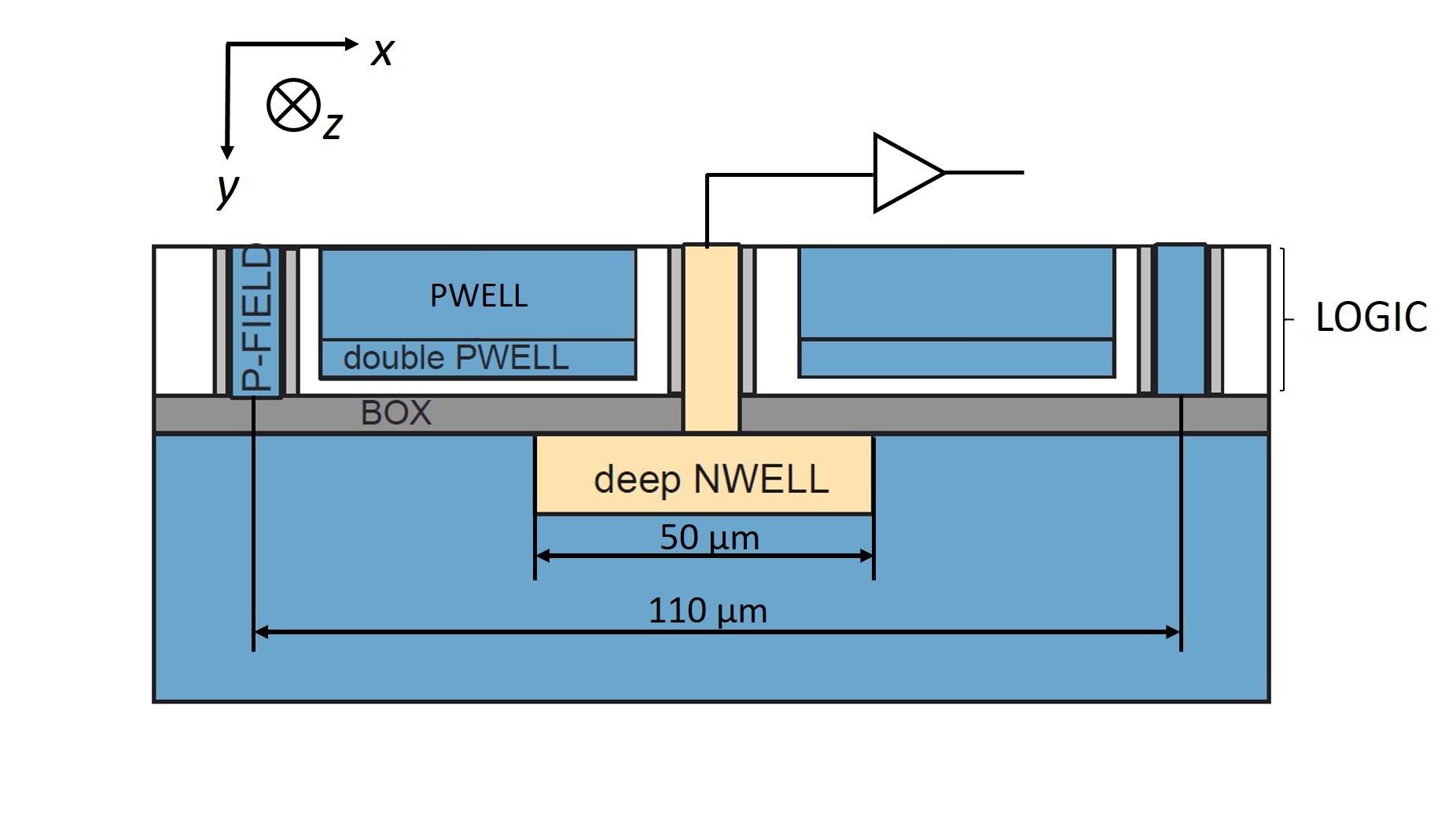}%
\caption{}
\label{fig:design}
\end{subfigure}
\caption{
(a) Microscope image of the array 2A on the XTB02 chip with indicated pads for connecting the single and peripheral pixels and the beam direction in E-TCT.
(b) Cross section of a pixel in the structure 2A. Shown are the deep n-well collecting electrode with the corresponding dimensions, the buried oxide (BOX), the LOGIC section on the p-type epitaxial layer (white) with p-wells for the active circuitry, and the P-FIELD implant for inter-pixel insulation. XTB02 is read out by connecting the amplifier directly to the deep n-well (figure taken from \cite{FernandezJinst}). The coordinate system is the one used in E-TCT measurements. Beam direction is along the $z$-axis. }%
\label{fig:sample}%
\end{figure}

Monolithic particle detectors have been used already for several years \cite{turchetta, Star, alice} but they were not suitable for use in the ATLAS experiment at the LHC because of their relatively slow speed and insufficient radiation hardness, since the dominant charge collection mechanism in these detectors is diffusion \cite{Deveaux-radhard}. Recently new technologies were developed permitting usage of high voltages and full CMOS circuitry on the same chip. This opened the possibility of designing active pixel detectors with sufficient depleted thickness for fast collection of charge drifting in the electric field \cite{PericHVCMOS}. Several other developments followed \cite{Havranek, HemperekNIM, Kishishita} commonly referred to as \emph{depleted CMOS pixels} \cite{Wermes-article}. Prototype test structures from various producers were tested recently and their performance after irradiation was investigated \cite{Affolder, AMSRistic, AMSLiu, AMSVitaliy, ChessRemoval}. 
One very interesting technological option is Silicon On Insulator (SOI), where Buried OXide (BOX) isolates the bulk from
the top layer where electronic circuitry exploiting full CMOS possibilities can be implemented. Since BOX is protecting
the electronics, high voltage can be applied to form a significant depleted layer in the bulk for fast charge collection.
A schematic drawing of a pixel investigated in this work is shown in Figure \ref{fig:design}.

Detector prototypes designed by the University of Bonn were produced in $180\,\mathrm{nm}$ SOI CMOS in the XFAB process \cite{XFab-process}. XFAB uses the thick film SOI process where double well structures shield FET transistors from the charge trapped in the BOX after irradiation. For this reason it is immune to the so called back gate effect \cite{back-gate} and can, unlike other SOI detectors, withstand high ionizing doses of over $700\,\mathrm{Mrad}$ \cite{FernandezNIM}.
Samples produced by XFAB were also tested with radioactive sources \cite{FernandezJinst} and in a test-beam experiment \cite{FernandezTB}
with good results before irradiation. 
In this paper results of Edge Transient Current Technique (E-TCT) measurements with XFAB test structures irradiated up
to $1\times 10^{16}\,\mathrm{n}_{\mathrm{eq}}/\mathrm{cm}^2$ are presented.

\section{Sample and experimental technique}
\label{sec:sample}

The chip investigated in this work is called XTB02. The substrate is p-type silicon with a resistivity of $\approx 100\,\Omega\cdot\mathrm{cm}$. The chip thickness is
$700\,\upmu\mathrm{m}$.
The XTB02 chip houses several test structures with different design parameters \cite{HemperekNIM, FernandezJinst}. 
The focus of this study is the test structure 2A, a $4\times 4$ pixel matrix with pixel dimension of $110\times 100\,\upmu\mathrm{m}^2$ shown in Figure \ref{fig:photo-array}.
Cross section of a pixel can be seen in Figure \ref{fig:design}.
The charge is collected by deep n-well electrodes with dimensions $40\times 50\,\upmu\mathrm{m}^2$ placed under the BOX layer. 
Above the BOX there is a few $\upmu\mathrm{m}$ thick epitaxial layer. In the p-well area above each deep n-well, called LOGIC, active CMOS circuitry can be placed. 
The epitaxial layer also contains an implant structure called P-FIELD around individual pixels. Its aim is to modify the electrical field bellow the BOX to break the conductive channel formed in the bulk due to BOX space charge formed after irradiation \cite{FernandezNIM,FernandezJinst}.
XTB02 devices are dedicated for studying the properties of the silicon bulk, so there is no readout circuitry actually implemented in the LOGIC. The charge collection electrode is connected directly to the external amplifier as shown in Figure \ref{fig:design}.
The substrate is biased via the outermost guard ring, a p-type implant ring surrounding the test structures. LOGIC and P-FIELD can be connected via separate bond pads. 
The deep n-well of one of the four central pixels is routed to an independent bonding pad while the other pixels are connected together and routed to another pad (seen in Figure \ref{fig:photo-array}).
This layout allows measurements on either a single pixel or on the entire array of 16 pixels, depending on which pad is connected to the amplifier.
In measurements described here the sample was always biased with positive high voltage applied to the deep n-wells while the substrate was kept at ground potential.
Unless otherwise noted, LOGIC was biased to the potential of the deep n-well and P-FIELD was left floating. The maximal allowable bias voltage was $300\,\mathrm{V}$. 

\begin{figure}%
\centering
\begin{subfigure}[b]{0.6\textwidth}
\centering
\includegraphics[width=\columnwidth]{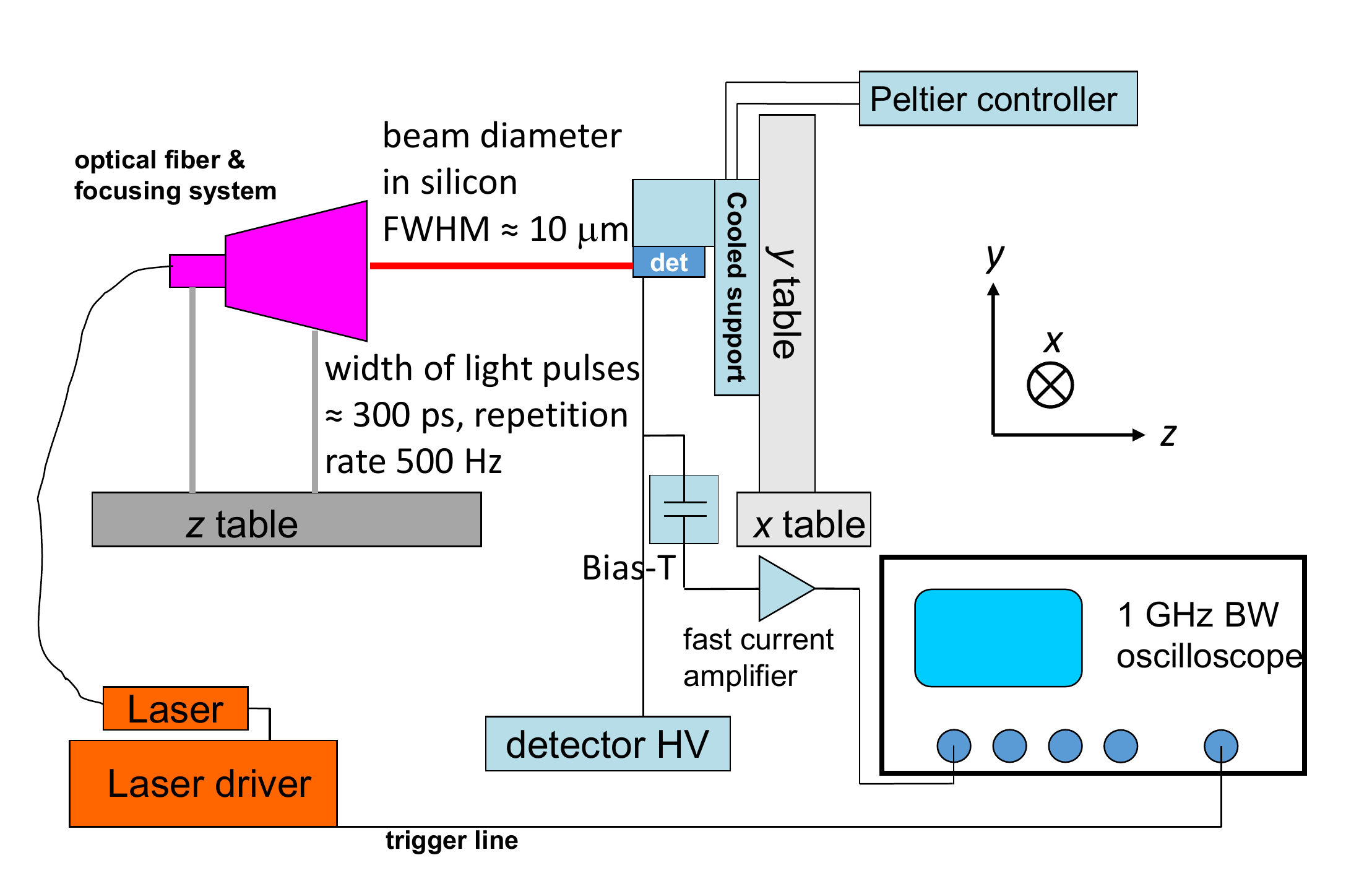}
\caption{}
\label{fig:TCT-setup-1}
\end{subfigure}%
\begin{subfigure}[b]{0.43\textwidth}
\centering
\includegraphics[width=\columnwidth]{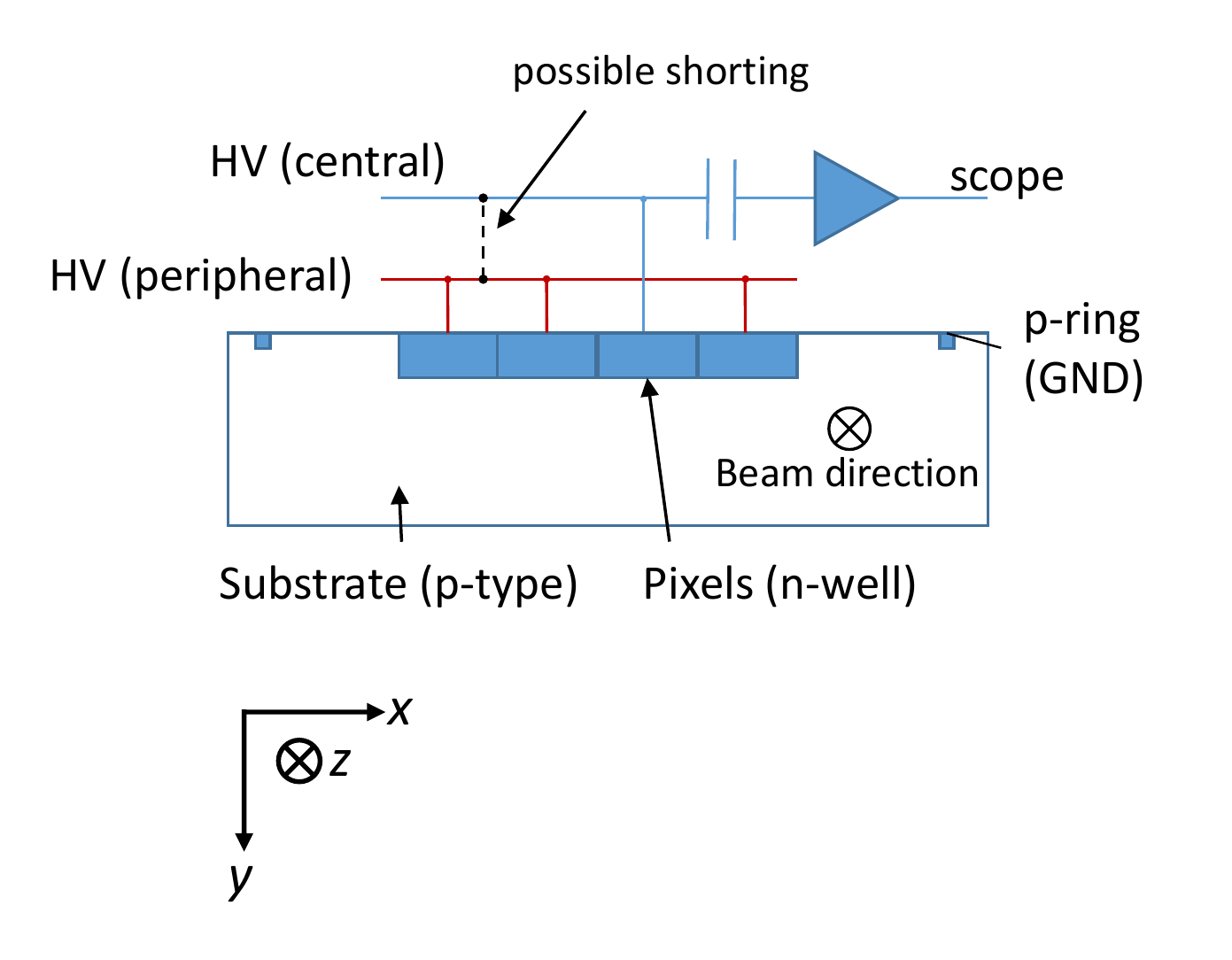}
\caption{}
\label{fig:TCT-setup-2}
\end{subfigure}%
\caption{Schematics of the E-TCT measurement setup (a) and the detector connection scheme (b).}%
\label{fig:TCT-setup}%
\end{figure}

The schematics of the system for E-TCT measurements and the detector connection are presented in Figure \ref{fig:TCT-setup}. The sample is placed with its edge in a focused beam of a pulsing infrared laser ($\lambda = 1064\,\mathrm{nm}$ -- absorption length in Si $\approx 1\,\mathrm{mm}$, pulse width $\leq 300\,\mathrm{ps}$, repetition rate $500\,\mathrm{Hz}$, FWHM of beam profile in focal point $\leq 10\,\upmu\mathrm{m}$). The light pulses generate electron-hole pairs along the beam path in the sample. The amount of generated charge is not calibrated, but the laser pulse power is kept constant to $\approx 5\,\%$ during individual measurements. 
Between different measurements the laser power was varied to some extent and different laser diodes were used, therefore the amount of injected charge cannot be directly compared between separate runs.
The position of the sample in the laser beam is controlled by a set of positioning stages with sub-$\upmu\mathrm{m}$ precision, 
which allow beam positioning in the $xy$-plane with a precision better than the beam width. Due to the large absorption length of the infrared light the deposited charge is roughly constant along the $z$-direction, meaning that the measurement has no resolution in this direction.
The charge carriers generated by the laser pulse start to drift in electric field, inducing an electric current on the readout electrodes. This current is amplified by a $1\,\mathrm{GHz}$ bandwidth current amplifier and digitized by a $1\,\mathrm{GHz}$ bandwidth oscilloscope. Waveforms of 50 pulses are averaged by the scope and stored to the computer. 
All measurements are carried out at room temperature, since there were no occurrences of the thermal runaway of the sensor at any irradiation stage.
A detailed description of the E-TCT method is available in \cite{TCT-setup}. Measurements reported here were made with an E-TCT system produced by Particulars \cite{particulars}.

The sample was irradiated with neutrons at Jo\v{z}ef Stefan Institute's research reactor \cite{Reactor}. A single sample was available for the study and it was therefore irradiated in steps. The cumulative fluence after each step was $2\times 10^{14}$, $5\times 10^{14}$, $1\times 10^{15}$, $2\times 10^{15}$, $5\times 10^{15}$ and $1\times 10^{16}\,\mathrm{n}_\mathrm{eq}/\mathrm{cm}^2$ with a $10\,\%$ error margin. 
During irradiation with neutrons samples also receive an ionizing dose of about $1\,\mathrm{kGy}$ (in $\mathrm{SiO}_2$) per each $1\mathrm{e}14\,\mathrm{n_{eq}}/\mathrm{cm}^2$. The dose was estimated using RadFETs, dedicated pMOS transistors, where TID is estimated from the change of threshold voltage \cite{Reactor-RADFET}. 
After each irradiation step the sample was annealed ($80\,\mathrm{min}$ at $60^\circ\,\mathrm{C}$) before E-TCT measurements were made. The sample was kept in the freezer at $-17^\circ\,\mathrm{C}$ and it was warmed to room temperature only for few hours during measurements. 

\section{Evolution of sensitive depth with irradiation}
\label{sec:depletion}

\begin{figure}%
\centering
\includegraphics[width=0.8\columnwidth]{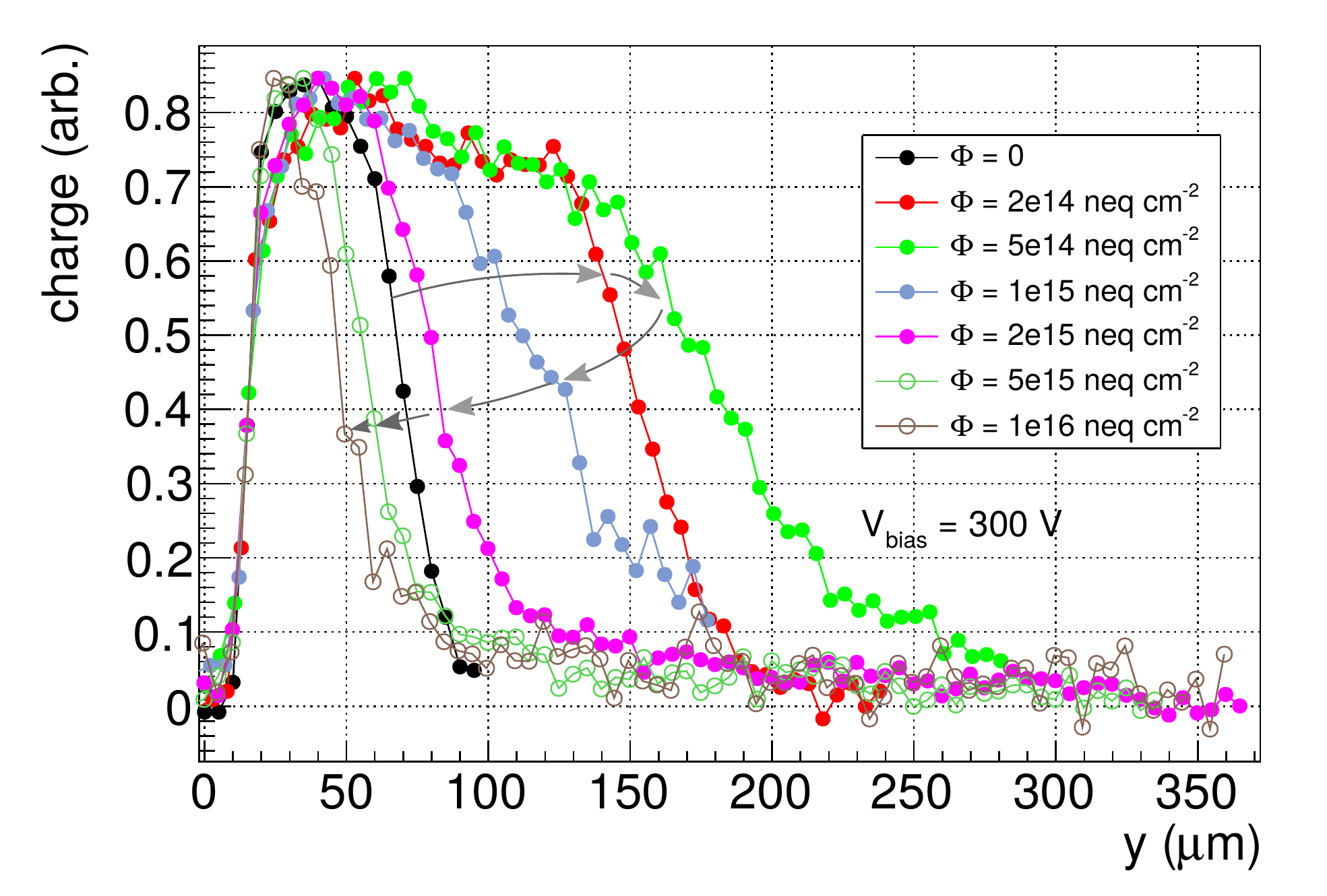}%
\caption{Normalized charge collection profiles along the pixel centre at $300\,\mathrm{V}$ bias for different neutron fluences. Arrows indicate the sequence of irradiation steps.}%
\label{fig:profile}%
\end{figure}

The collected charge in E-TCT is defined as the integral of the induced current pulse over $25\,\mathrm{ns}$ after the beginning of the pulse.  
The collected charge was measured with a single pixel connected to readout. Laser was directed to different depths $y$ (see Figure \ref{fig:TCT-setup} for definition of the coordinate system), while the horizontal beam position $x$ was fixed to the centre of a pixel.
The measured charge as a function of the beam position is called charge collection profile. Charge collection profiles measured before irradiation and after each fluence step are shown in Figure \ref{fig:profile}. Measurements were taken at the highest bias voltage of $300\,\mathrm{V}$. The curves were normalized to the same maximal value. 
The transition at the rising edge of the charge collection profile (charge close to the chip surface, at $y \approx 20\,\upmu\mathrm{m}$) corresponds to the laser beam gradually entering the sample. The transition takes place over $\approx 10\,\upmu\mathrm{m}$ which corresponds to the laser beam diameter. 
Once the transition at the chip surface is finished, the beam is for a while fully contained within the depletion zone. 
This results in a plateau in the charge collection profile. The slope of the top of the profile seen in measurements after irradiation may be related to charge trapping as the electric field and thus the carrier velocity is falling with distance from the surface. 
The transition on the falling edge of the charge collection profile is slower than on the surface side --- the collected charge typically reduces from $90\,\%$ to $10\,\%$ of the maximum value over a depth of approximately $50\,\upmu\mathrm{m}$, --- which is due to the form of the depletion zone deviating from the abrupt junction approximation.
Figure \ref{fig:profile} shows that the width of the charge collection profile increases with irradiation, reaching a maximum at a fluence of $5\times 10^{14}\,\mathrm{n}_\mathrm{eq}/\mathrm{cm}^2$. The width reduces at higher fluences, however it is still significant even at the highest fluence of $1\times 10^{16}\,\mathrm{n}_\mathrm{eq}/\mathrm{cm}^2$.
This behaviour is consistent with radiation induced removal of initial acceptors \cite{ChessRemoval}. Upon first irradiation steps the effective space charge concentration is reduced as more initial acceptors are removed than new are created by irradiation. After the removal process is finished the negative space charge concentration increases with increasing fluence.

\begin{figure}%
\centering
\includegraphics[width=0.8\columnwidth,trim=0 0 0 0.07\columnwidth,clip]{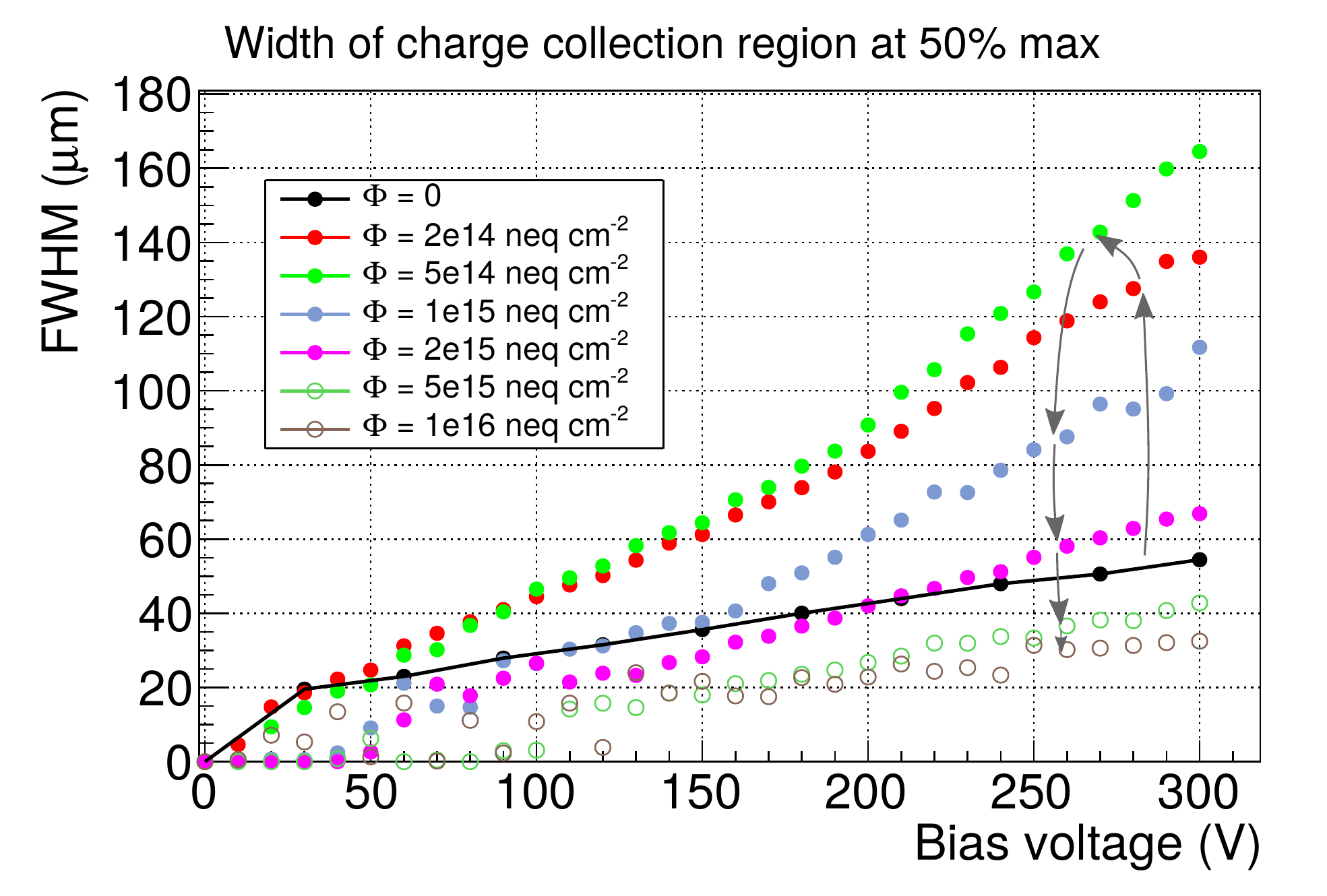}%
\caption{Width of charge collection profiles vs. bias voltage for different neutron fluences. Arrows indicate the sequence of irradiation steps.}%
\label{fig:depletion}%
\end{figure}

The width of the sensitive region was quantified by evaluating the full width at half maximum of the charge collection profiles. 
Charge collection width (i.e. profile width) as a function of bias voltage is shown in Figure \ref{fig:depletion} for all fluences.
The width of the depletion region in a sensor with planar electrode geometry is calculated in the model of constant space charge as $W_\mathrm{depl} = \sqrt{\frac{2\epsilon}{q_0 N_\mathrm{eff}} V}$, with $\epsilon$ the electric permittivity of the bulk, $q_0$ the elementary charge, $N_\mathrm{eff}$ the effective acceptor concentration in the space charge region and $V$ the applied bias voltage. 
With the rest of the quantities known, one can use this dependence to extract the values of $N_{\mathrm{eff}}$ for different fluences.
For the measurement before irradiation a fit could be made, yielding the value of $N_\mathrm{eff} \approx 1.3\times 10^{14}\,\mathrm{cm}^{-3}$, consistent with the initial resistivity of the substrate.
After irradiation the width of the sensitive region grows faster than the square root function of voltage and the fit is therefore not possible.
%
In Figure \ref{fig:depletion} an unusual behaviour at $1\times 10^{15}\,\mathrm{n}_\mathrm{eq}/\mathrm{cm}^2$ and higher fluences can be observed at low bias voltages, where one can notice very low charge collection width for the initial bias voltages followed by a relatively fast increase, which looks like a certain threshold voltage has to be reached before charge collection starts. The mechanism behind this behaviour is not understood. 

\section{Effective acceptor removal parameters}

\begin{figure}%
\centering
\includegraphics[width=0.7\columnwidth]{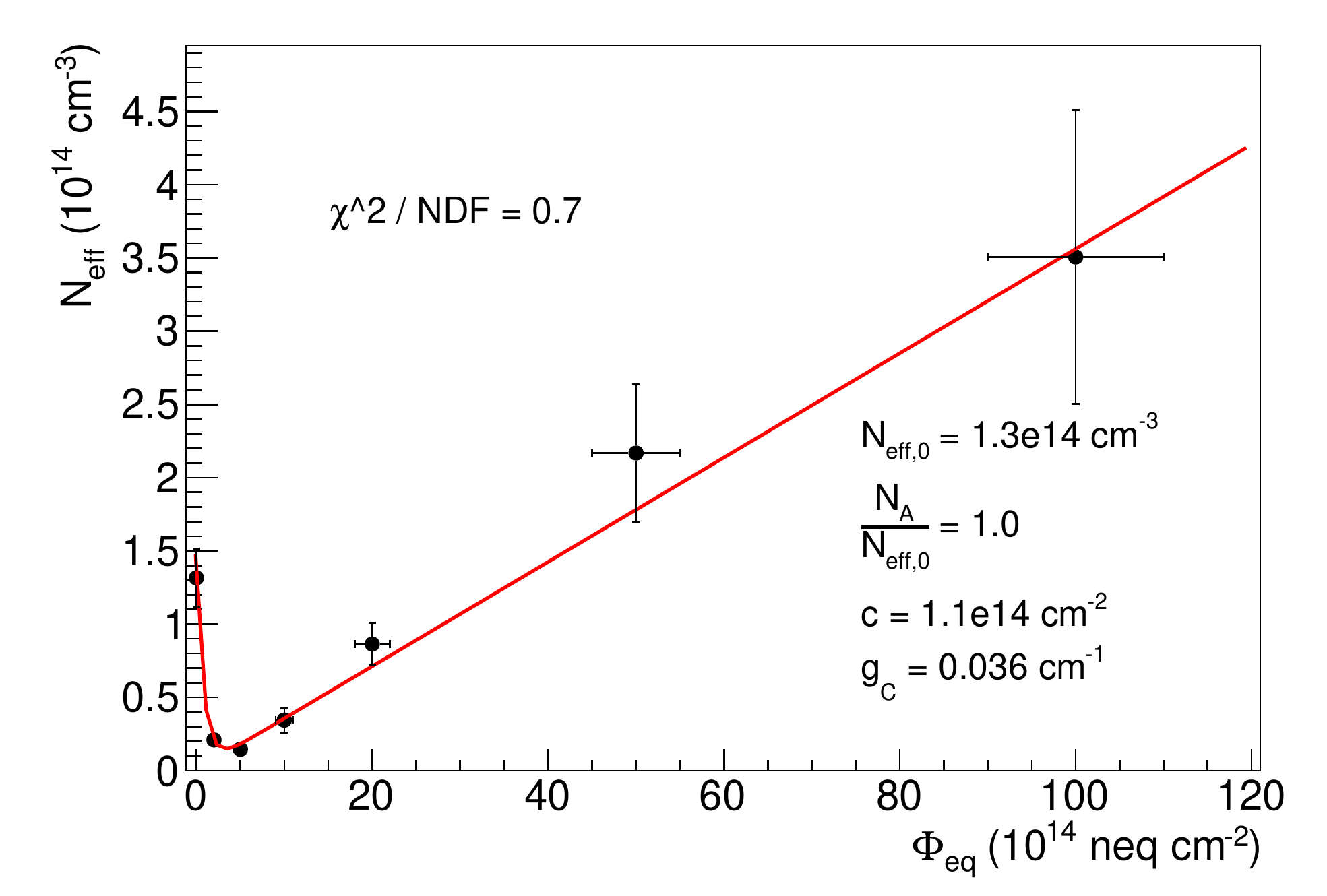}%
\caption{Evolution of $N_{\mathrm{eff}}$ of the substrate with fluence. The fit of the data with the function \ref{eq:Neff_vs_fluence} and the extracted fit parameters are also shown.}%
\label{fig:neff}%
\end{figure}

Irradiation of silicon with neutrons introduces defects into its crystal structure. Interaction of initial dopants with these defects may turn them electrically neutral so that they do not contribute to the effective space charge any more \cite{Wunstorf, RD48}.
The radiation induced defects also form negatively charged localized energy levels which contribute to the effective acceptor concentration in the depleted layer. If the number of neutralized initial acceptors is larger than the number of newly created acceptors the effective space charge decreases, therefore increasing depleted depth.
The increase of depleted depth with irradiation was clearly observed as shown in Figure \ref{fig:profile} and the behaviour can be explained by the removal of initial acceptors. 
This has a beneficial impact on the signal after irradiation.
As mentioned in the previous section the dependence of depleted depth on bias voltage does not follow the simple square root behavior after irradiation. However, to get some comparison of acceptor removal behavior with other measurements \cite{ChessRemoval, MandicLF, Cavallaro} the following procedure was made. $N_\mathrm{eff}$ was calculated from the width of the charge collection profiles by evaluating the formula $W_\mathrm{depl} = \sqrt{\frac{2\epsilon}{q_0 N_\mathrm{eff}} V}$ at the bias voltage of $300\,\mathrm{V}$. A systematic uncertainty on the value of $N_\mathrm{eff}$ was estimated from the spread of the values when taking the width of charge collection profiles at $(50\pm10)\,\%$ of the maximum.
The measured values of $N_{\mathrm{eff}}$ at different fluences calculated at a bias voltage of $300\,\mathrm{V}$ are shown in Figure \ref{fig:neff}. 
The evolution of $N_\mathrm{eff}$ as a function of fluence is given by \cite{Affolder, ChessRemoval, RD48}: 
\begin{equation}
N_\mathrm{eff} = N_{\mathrm{eff},0} - N_\mathrm{A} (1-\exp(-c\cdot \Phi_\mathrm{eq})) + g_\mathrm{C}\Phi_\mathrm{eq},
\label{eq:Neff_vs_fluence}
\end{equation} 
where $N_{\mathrm{eff},0}$ denotes the initial effective acceptor concentration of the substrate, $N_{\mathrm{A}}$ the concentration of the effectively removed acceptors, 
$c$ the acceptor removal constant, $\Phi_\mathrm{eq}$ the $1\,\mathrm{MeV}$ neutron equivalent fluence and $g_{\mathrm{C}}$ the generation rate of stable deep acceptors \cite{Cindro}.
The measured data were fit with function \ref{eq:Neff_vs_fluence} with $N_{\mathrm{eff},0}$, $N_{\mathrm{A}}$, $c$ and $g_\mathrm{C}$ as free parameters. Results of the fit are shown in Figure \ref{fig:neff}. 
The ratio $N_{\mathrm{A}} / N_{\mathrm{eff},0} = 1$ indicates a complete initial acceptor removal.
The value of the parameter $c = 1.1\times 10^{-14}\,\mathrm{cm}^2$ is consistent with that in \cite{Cavallaro} measured on a substrate of the same initial resistivity. 
At the same time it is by a factor of 2--3 larger than for substrates of initial resistivities of 10 and $20\,\Omega\cdot\mathrm{cm}$ measured in \cite{ChessRemoval}. This is consistent with the observation that the acceptor removal constant is smaller in silicon with a lower initial resistivity \cite{Krambi-LGAD}.
The value of $c$ is also reflected in the fluence at which the maximal charge collection width is reached. In the XTB02 sample the maximum is reached at $\sim 5\times 10^{14}\,\mathrm{n}_\mathrm{eq}/\mathrm{cm}^2$ whereas for samples in \cite{ChessRemoval} it occurs at $\sim 2\times 10^{15}\,\mathrm{n}_\mathrm{eq}/\mathrm{cm}^2$. 
The value of the parameter $g_\mathrm{C} = 0.036\,\mathrm{cm}^{-1}$ is larger than the value $0.02\,\mathrm{cm}^{-1}$ usually observed for neutron irradiated samples \cite{ChessRemoval}.
But this is not surprising since we know that the depleted depth does not follow the $\sqrt{V}$ behaviour, pointing to an inconsistency with the uniform space charge concentration and abrupt junction approximation.  
By evaluating $N_\mathrm{eff}$ from measurements at a bias voltage of $210\,\mathrm{V}$ we for example obtain the value of $g_\mathrm{C}=0.08\,\mathrm{cm}^{-1}$, whereas at $150\,\mathrm{V}$ the value is $g_\mathrm{C}=0.2\,\mathrm{cm}^{-1}$. However, the other three fit parameters are stable within $10\,\%$ of the value at $300\,\mathrm{V}$, because they are related to the maximum of depleted width, which is at $\sim 5\times 10^{14}\,\mathrm{n}_\mathrm{eq}/\mathrm{cm}^2$ at all bias voltages.
This rises the confidence that the value of the acceptor removal constant extracted from the fit in Figure \ref{fig:neff} is a good estimate for this substrate material. 
 
\section{Uniformity of charge collection efficiency}
\label{sec:efficiency}
\begin{figure}%
\centering
\includegraphics[width=0.6\columnwidth,trim=0 0 0 0.08\columnwidth,clip]{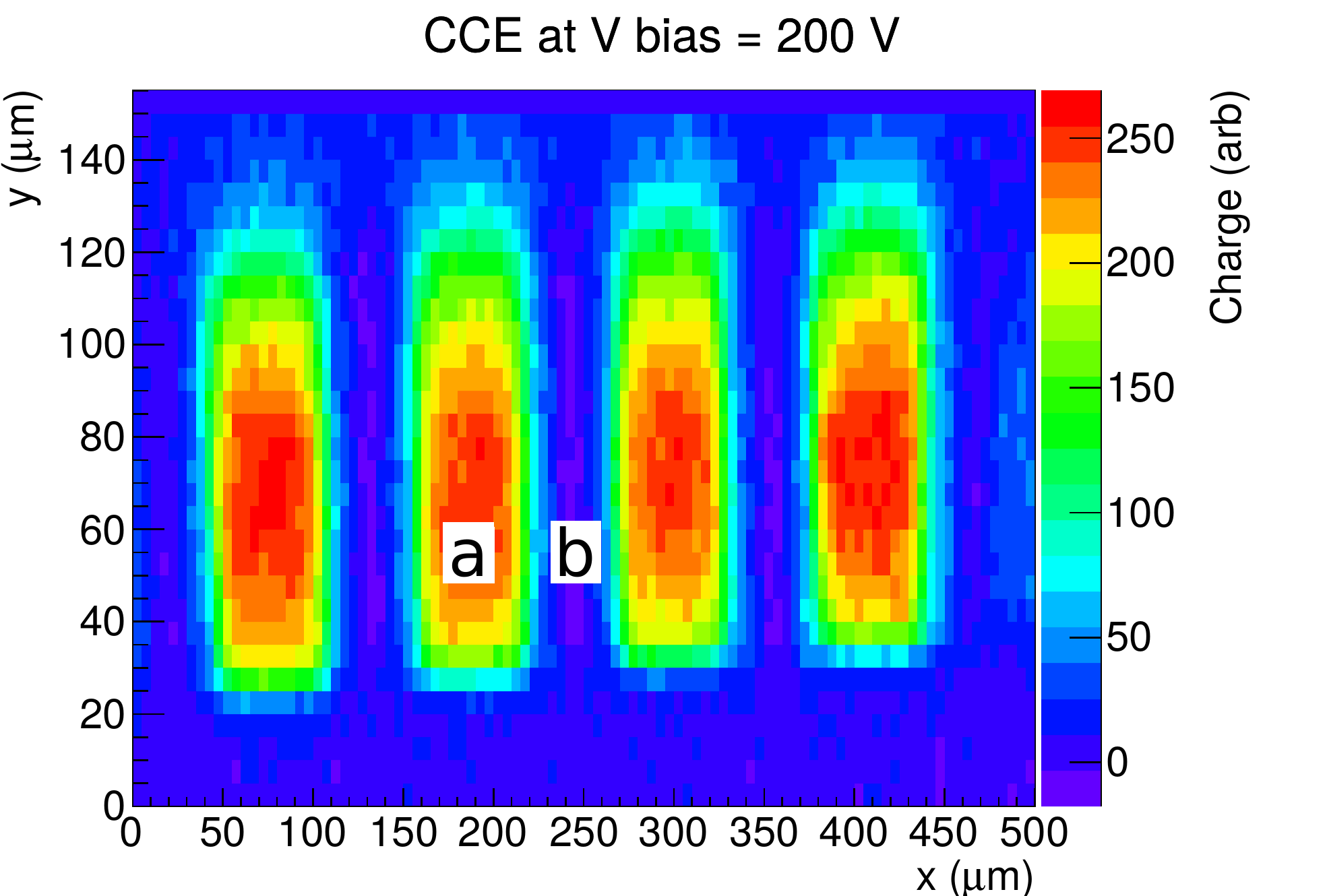} 
\vspace{0.3cm}
\\
\begin{subfigure}[t]{0.5\textwidth}
\centering
\includegraphics[width=0.9\columnwidth,trim=0 0 0 0.15\columnwidth,clip]{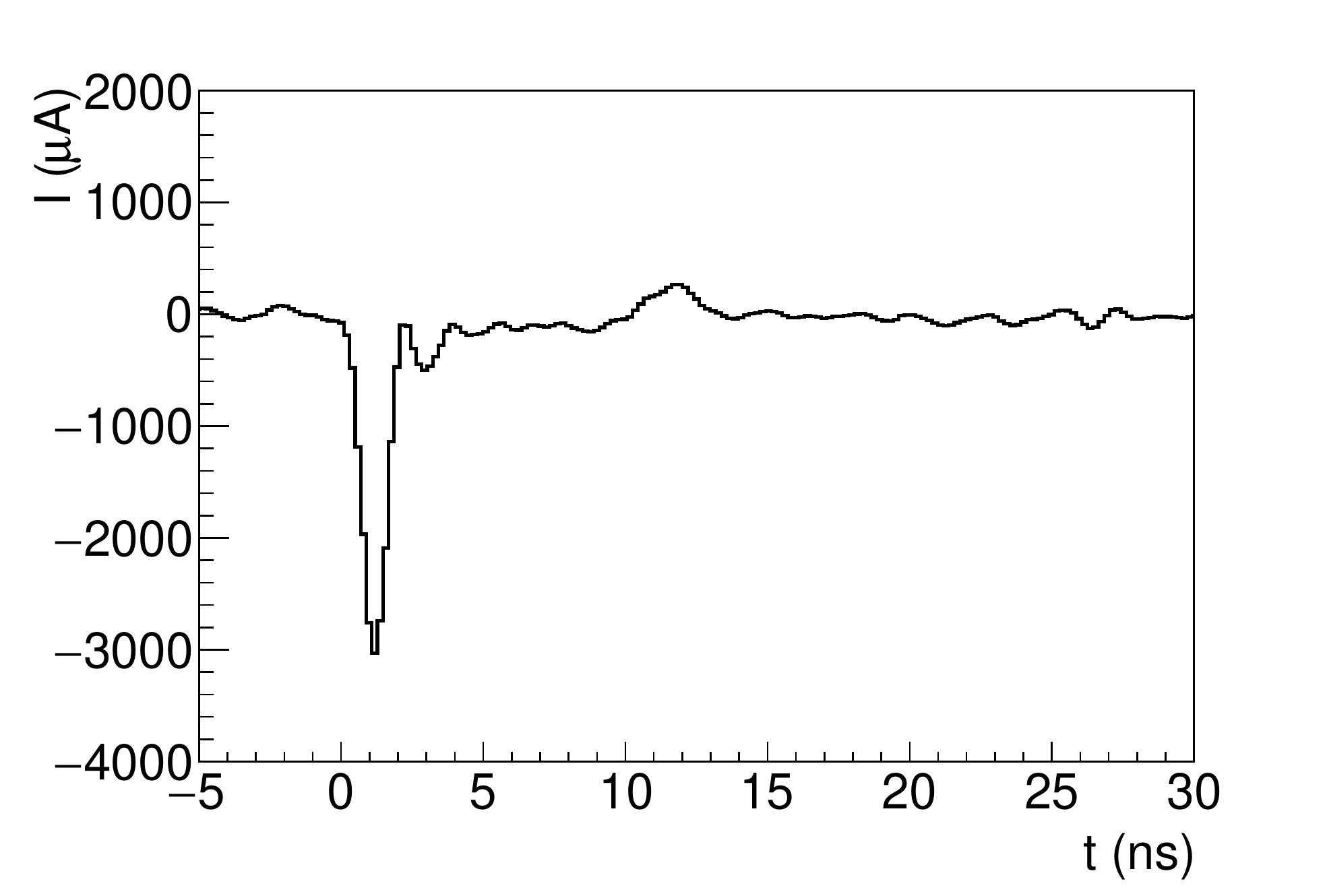}
\caption{High efficiency}
\label{fig:pulse_high_eff}
\end{subfigure}%
\begin{subfigure}[t]{0.5\textwidth}
\centering
\includegraphics[width=0.9\columnwidth,trim=0 0 0 0.15\columnwidth,clip]{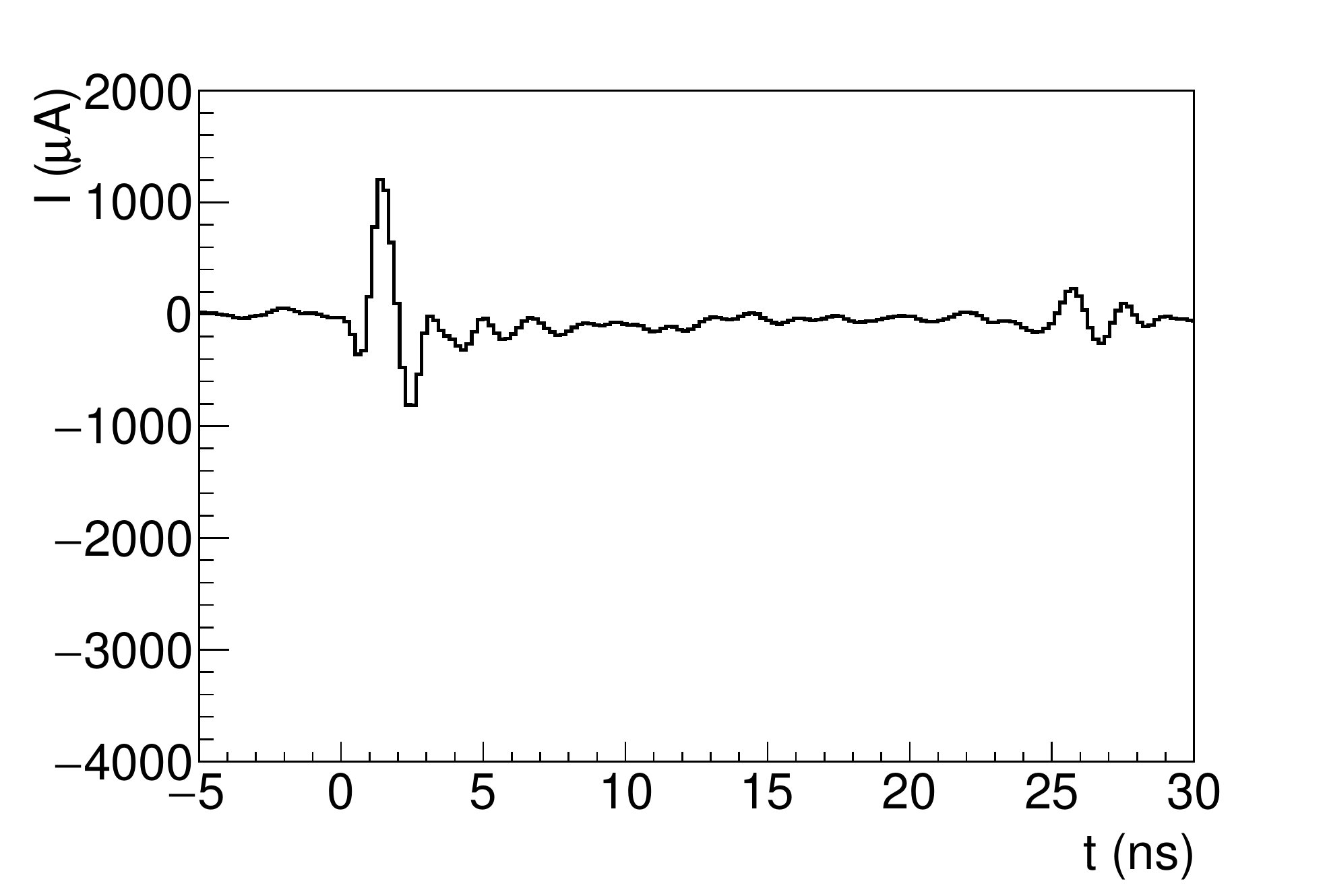}
\caption{Low efficiency}
\label{fig:pulse_low_eff}
\end{subfigure}%
\caption{Two dimensional charge collection profile in a detector irradiated to $5\times 10^{14}\,\mathrm{n}_\mathrm{eq}/\mathrm{cm}^2$ at $V_\mathrm{bias}=200\,\mathrm{V}$ with induced pulses in regions with high and low charge collection efficiency respectively.}%
\label{fig:gaps}%
\end{figure}

An important requirement for a pixel detector is the response uniformity within a pixel. The uniformity can be studied with a two dimensional E-TCT scan, where the position of the laser beam is varied along the sample depth (coordinate $y$) as well as along the edge of the sample (coordinate $x$, see Figure \ref{fig:TCT-setup}).
Figure \ref{fig:gaps} shows collected charge as a function of coordinates $x$ and $y$ after a fluence of $5\times 10^{14}\,\mathrm{n}_\mathrm{eq}/\mathrm{cm}^2$.
In the measurement all sixteen pixels of the test structure are connected together and read out simultaneously. Each region with a high collected charge corresponds to a column of four pixels along the $z$-axis, which cannot be distinguished from each other in an E-TCT measurement. 
Distinct regions with no collected charge can be seen between the columns. Size and spacing of these efficiency gaps roughly coincides with the spacing between deep n-wells of neighbouring pixels. As can be seen in Figure \ref{fig:design}, the n-well does not extend over the entire area of the pixel.

Figures \ref{fig:pulse_high_eff} and \ref{fig:pulse_low_eff} show the induced current pulses from an efficient and an inefficient region at $y=50\,\upmu\mathrm{m}$.
While the former is a unipolar pulse with a non-zero integral (with superimposed oscillation due to non-matching impedances of the cable and the amplifier), the latter is bipolar with a vanishing integral. 
A similar magnitude of both pulses confirms a uniform strength of electric field at a given sample depth.
According to the Shockley-Ramo theorem of signal formation, a bipolar pulse with zero integral is observed when the drift path of charge carriers does not end on a readout electrode \cite{shockley, ramo}.
The pulse \ref{fig:pulse_low_eff} therefore indicates a presence of a parasitic charge collecting electrode which is not connected to readout.
The identity of this electrode can be deduced from Figure \ref{fig:design}. It can be seen that there are two biased structures present on the top of the pixel --- the deep n-well, which is read out, and LOGIC, which is biased separately to the potential of the deep n-well but does not have a low impedance connection to the readout. Although LOGIC is positioned above the insulating BOX layer, it still influences the electric field in the bulk. If this effect is strong the field lines will be roughly perpendicular to the chip surface. When charge is not injected directly underneath the collecting n-wells, its drift path will end on the BOX rather than on the deep n-well, resulting in a low collected charge (Figure \ref{fig:sim_layout}). 
LOGIC therefore acts as an AC coupled parasitic electrode. This hypothesis was confirmed by switching the deep n-well and LOGIC connections, so that LOGIC was read out. This yielded a complementary picture --- low efficiency for charge injection underneath the deep n-wells and high efficiency for injection underneath LOGIC.

\begin{figure}%
\begin{subfigure}[t]{0.5\textwidth}
\includegraphics[width=0.9\columnwidth,trim=0 0 0 0.15\columnwidth,clip]{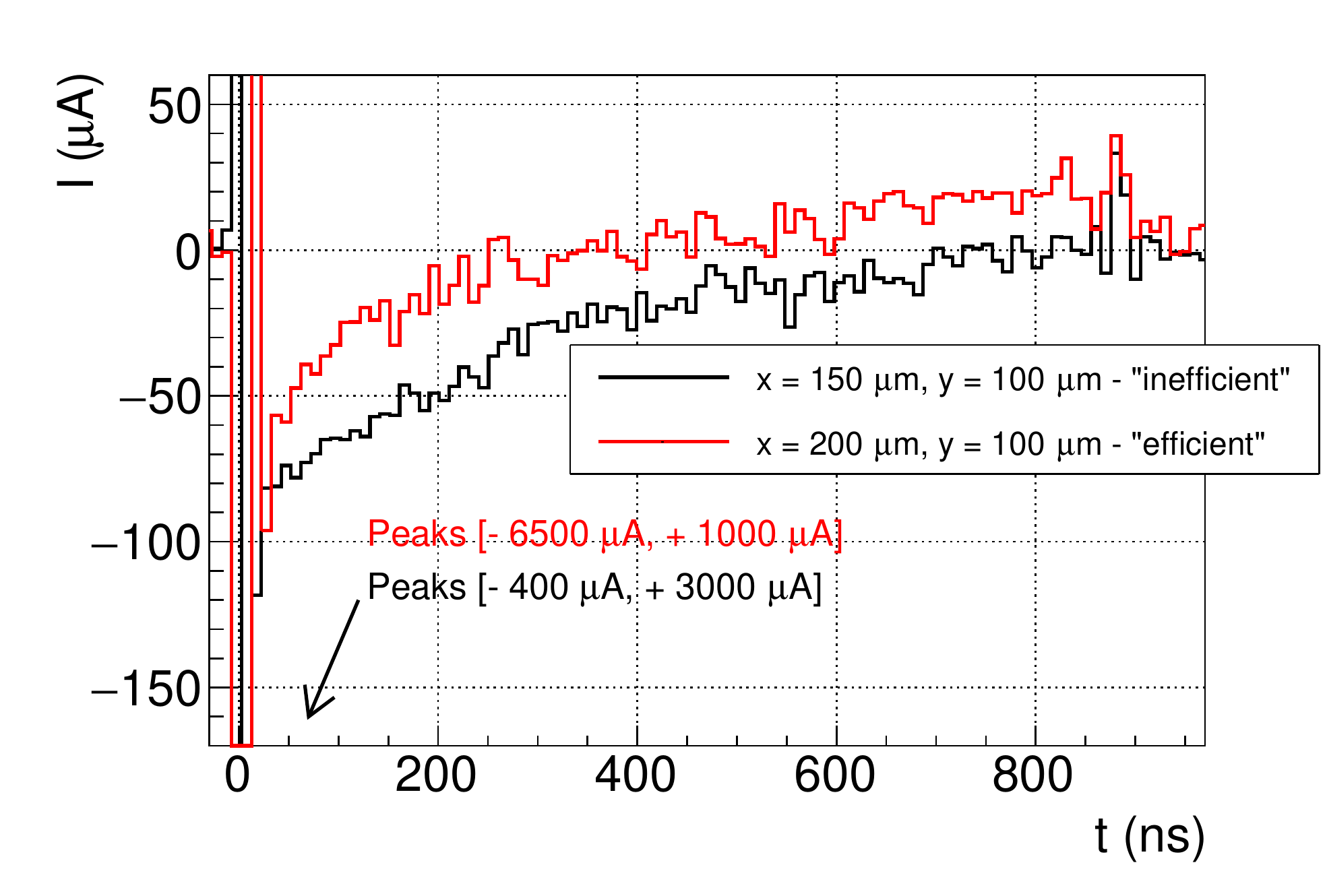}
\caption{}
\label{fig:long_pulse_a}
\end{subfigure}%
\hfill
\begin{subfigure}[t]{0.5\textwidth}
\centering
\includegraphics[width=0.9\columnwidth,trim=0 0 0 0.15\columnwidth,clip]{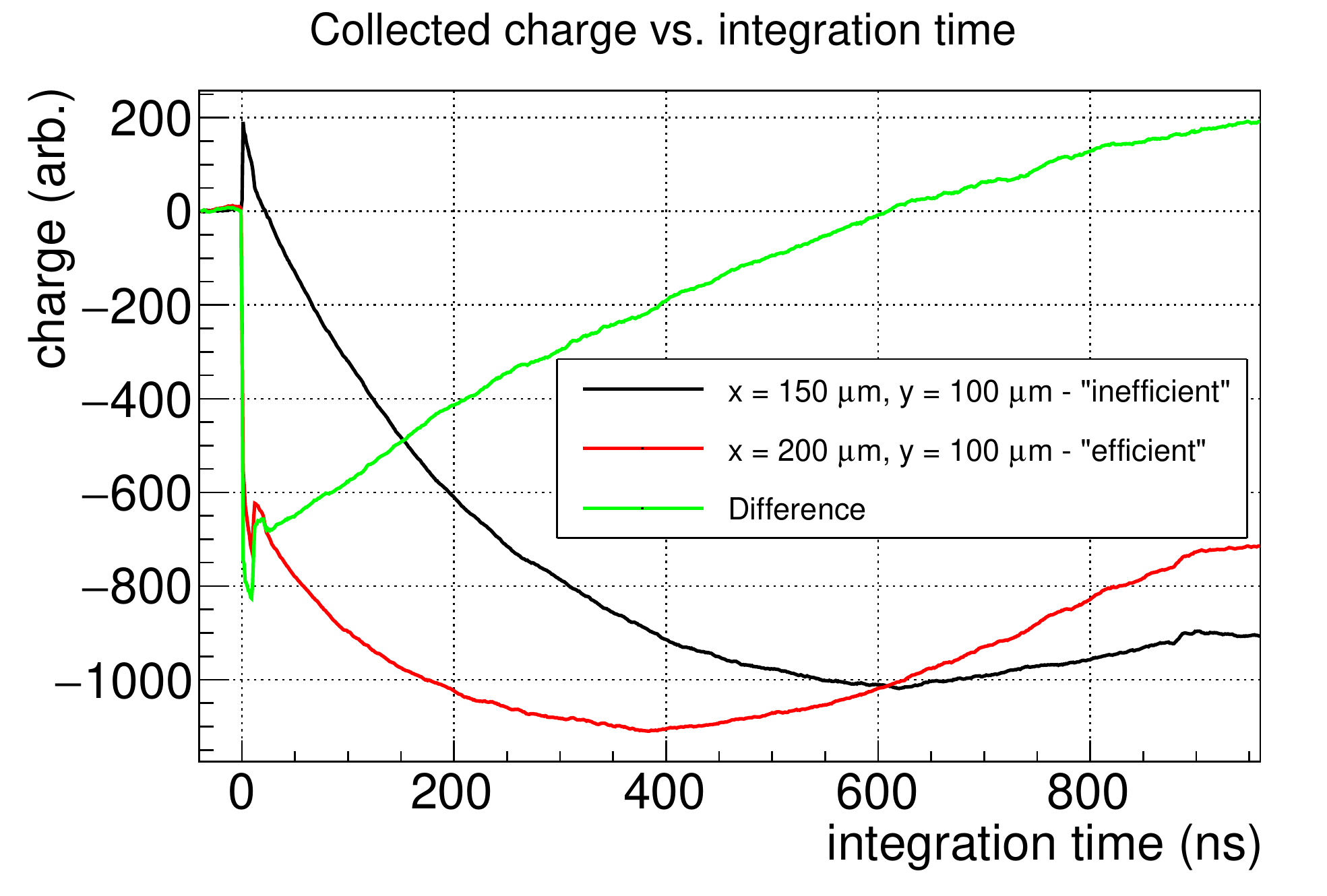}
\caption{}
\label{fig:long_pulse_b}
\end{subfigure}%
\\
\\
\\
\begin{subfigure}[t]{0.5\textwidth}
\centering
\includegraphics[width=0.9\columnwidth,trim=0 0 0 0.15\columnwidth,clip]{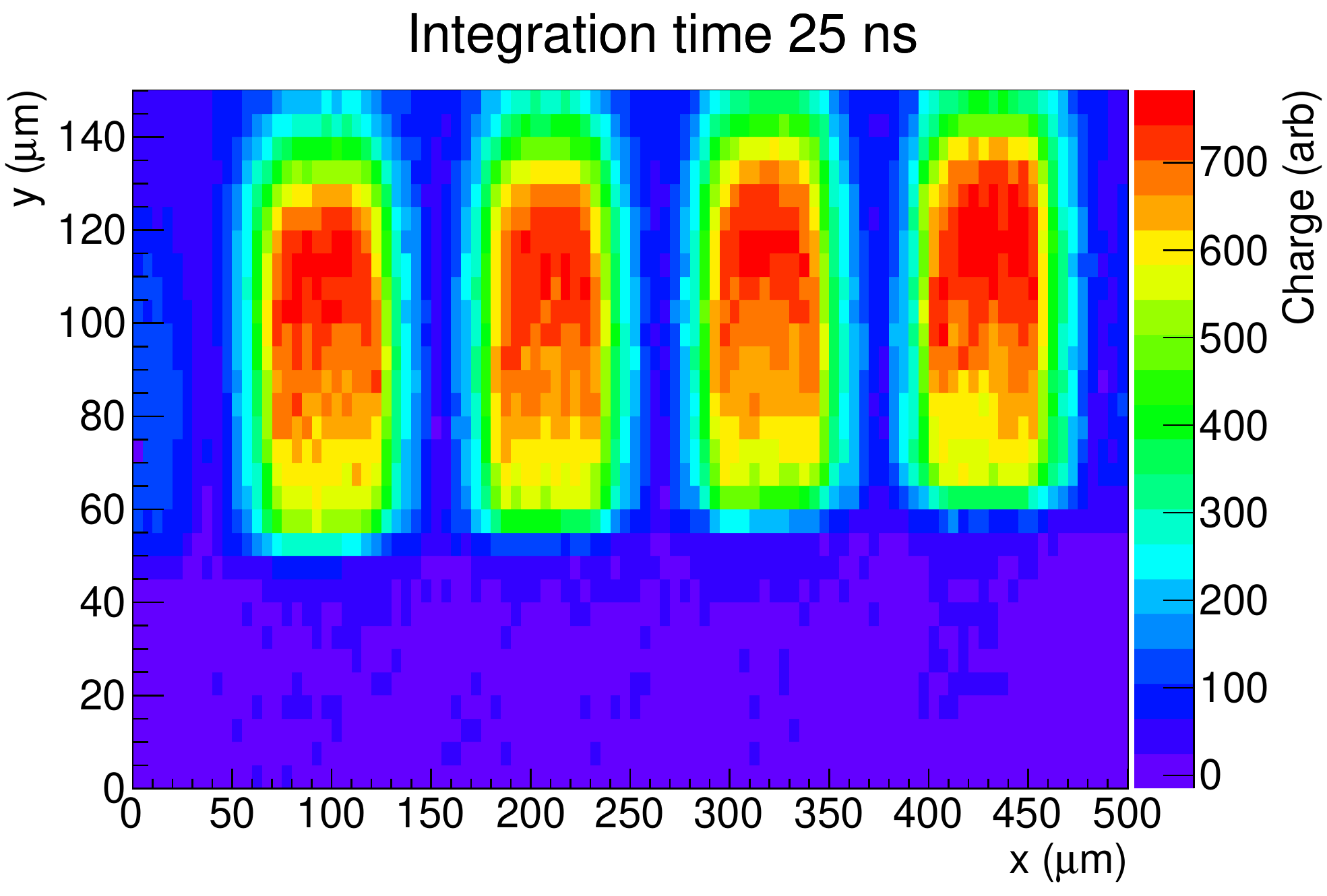}
\caption{Integration time $25\,\mathrm{ns}$}
\label{fig:tint_25ns}
\end{subfigure}%
\begin{subfigure}[t]{0.5\textwidth}
\centering
\includegraphics[width=0.9\columnwidth,trim=0 0 0 0.15\columnwidth,clip]{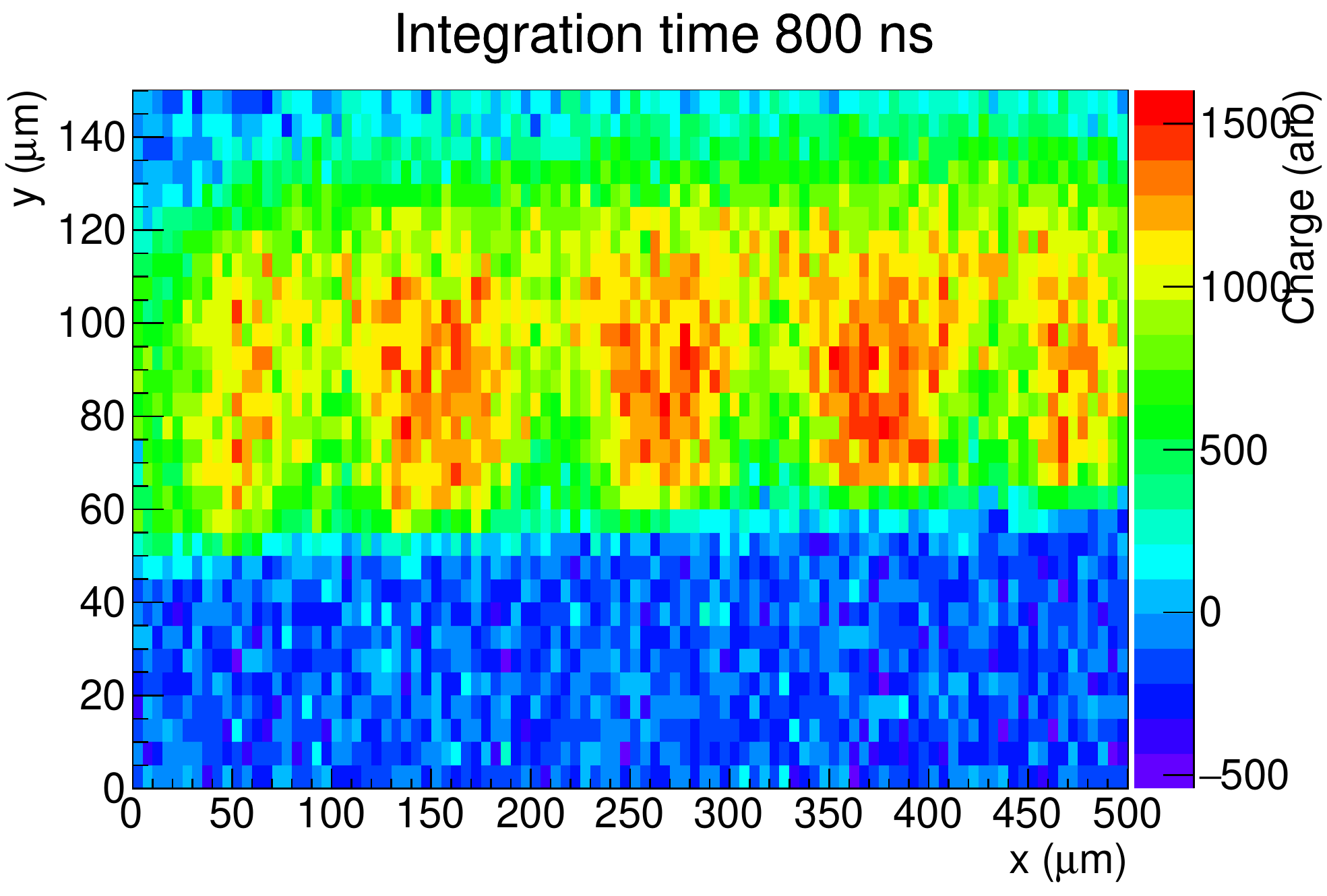}
\caption{Integration time $800\,\mathrm{ns}$}
\label{fig:tint_800ns}
\end{subfigure}%
\caption{(a) Induced electrical pulses from different efficiency regions on a time scale of $1\,\upmu\mathrm{s}$ in a detector irradiated to $5\times 10^{14}\,\mathrm{n}_\mathrm{eq}/\mathrm{cm}^2$. The drift contribution in the first $25\,\mathrm{ns}$ is truncated in the image, since it exceeds $1000\,\upmu\mathrm{A}$. The persisting pulse comes from the lateral current flow on the BOX interface. (b) Dependence of the pulse integral (collected charge) from integration time. (c) Charge collection profile with $25\,\mathrm{ns}$ integration time. (d) Charge collection profile with $800\,\mathrm{ns}$ integration time. The $z$-axis scales are the same as in figure \ref{fig:gaps}. The difference in measured charge comes from different laser power. All measurements were made at $200\,\mathrm{V}$ bias voltage.}%
\label{fig:long_pulse}%
\end{figure}

The electrons drifting towards LOGIC eventually stop on the non-conductive BOX layer after $\approx 10\,\mathrm{ns}$. Since charge cannot accumulate on BOX indefinitely, it has to be removed laterally towards the deep n-wells, which are the only conductive connection through the BOX. 
The lateral charge flow should result in an additional measurable induced current.
To investigate this assumption, the time scale of the E-TCT measurements was prolonged from $25\,\mathrm{ns}$ to $1\,\upmu\mathrm{s}$. Standard detector biasing scheme was used and the deep n-well was read out.
The long time scale pulses from an efficient and an inefficient region of the test structure are compared in Figure \ref{fig:long_pulse_a}. After the initial charge collection by drift in the first $25\,\mathrm{ns}$ is finished, a current pulse with a small amplitude ($<1\,\%$ amplitude from drift) persists for $1\,\upmu\mathrm{s}$. 
Note that on the scale used in Figure \ref{fig:long_pulse} the initial part of the pulse is truncated as it extends much below $-1000\,\upmu\mathrm{A}$.
The amplitude of this trailing pulse is higher in an inefficient region than in an efficient region. This current appears due to the charge collected on the BOX slowly discharging to the n-wells. Even in an efficient region some sections are not covered by n-wells due to the segmentation along $z$-direction, hence the current is always present. The exact mechanism of the current flow on the BOX surface was not addressed in this work.
The cumulative time integral of both pulses is shown in Figure \ref{fig:long_pulse_b}. Two characteristic regimes can be observed. For a short integration time of $25\,\mathrm{ns}$ about $\frac{2}{3}$ of the maximum charge is collected from an efficient region, whereas the charge collected from an inefficient region is zero, as already observed in Figure \ref{fig:gaps}.
However, for an integration time of several hundred $\mathrm{ns}$ the small difference in the amplitude of the pulse tail becomes important. At $600-800\,\mathrm{ns}$ integration time the collected charge from both regions is about the same, 
while for even longer integration times the collected charge in an inefficient region is even slightly higher than in an efficient region. 
The two dimensional charge collection profiles corresponding to $25\,\mathrm{ns}$ and $800\,\mathrm{ns}$ charge integration times are shown in Figures \ref{fig:tint_25ns} and \ref{fig:tint_800ns}. While efficiency gaps are present for the $25\,\mathrm{ns}$ integration time, the collected charge is relatively uniform for the $800\,\mathrm{ns}$ integration. 
These measurements are in agreement with the hypothesis of a parasitic charge collection via the AC coupled LOGIC electrode. The part of the charge which ends its drift on the BOX interface is effectively lost for $25\,\mathrm{ns}$ charge collection time. 
The charge from the BOX interface is transported to the collecting electrode by a much slower process, which takes about $1\,\upmu\mathrm{s}$ to complete as seen in Figure \ref{fig:long_pulse}.


\begin{figure}%
\centering
\begin{subfigure}[t]{0.47\textwidth}
\includegraphics[width=\columnwidth,trim=0 0 0 0.15\columnwidth,clip]{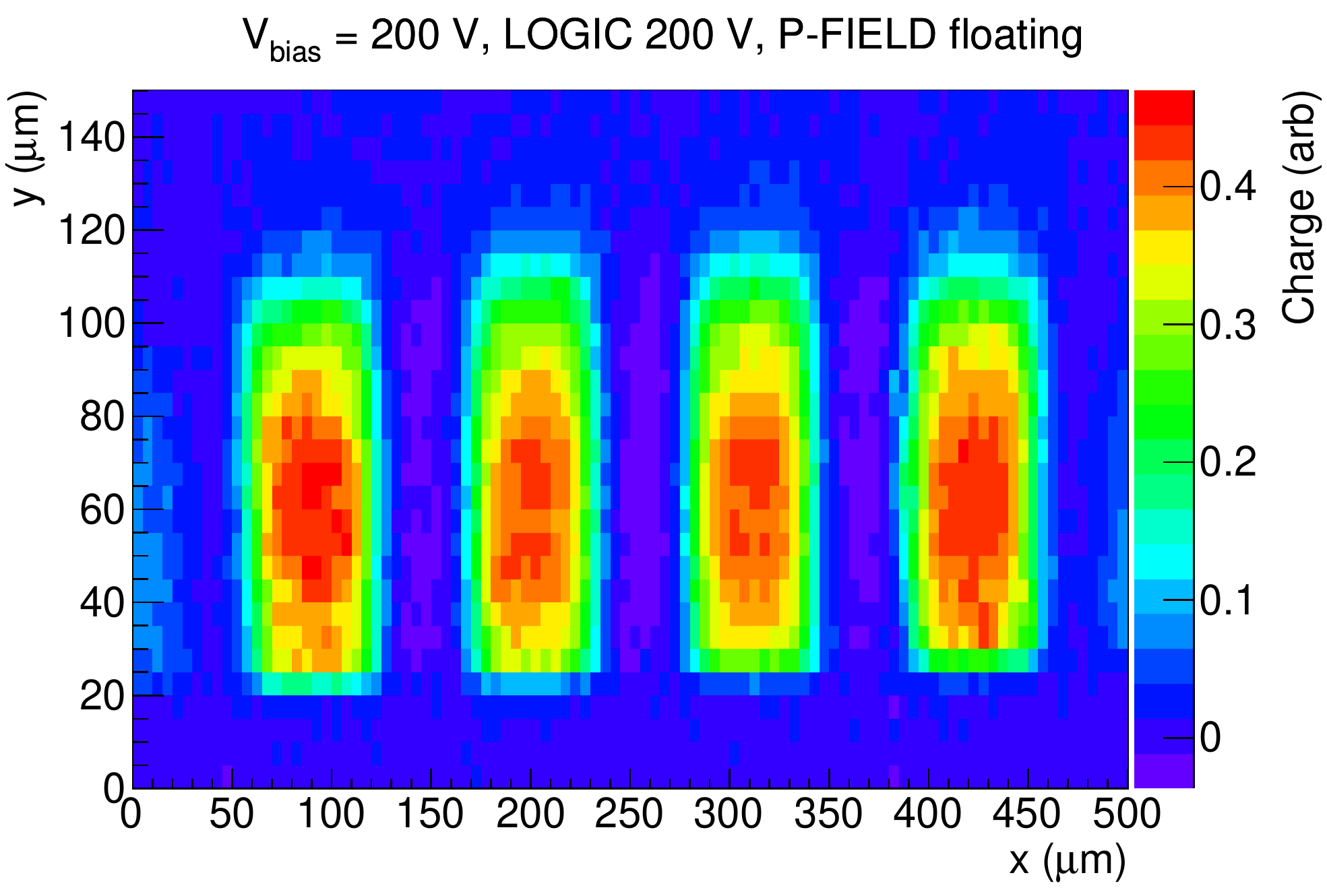}
\caption{$V_{\mathrm{bias}}=200\,\mathrm{V}$, $\mathrm{LOGIC}=200\,\mathrm{V}$}
\label{fig:gaps_200V}
\end{subfigure}%
\\
\vspace{0.4cm}
\begin{subfigure}[t]{0.47\textwidth}
\includegraphics[width=1\columnwidth,trim=0 0 0 0.15\columnwidth,clip]{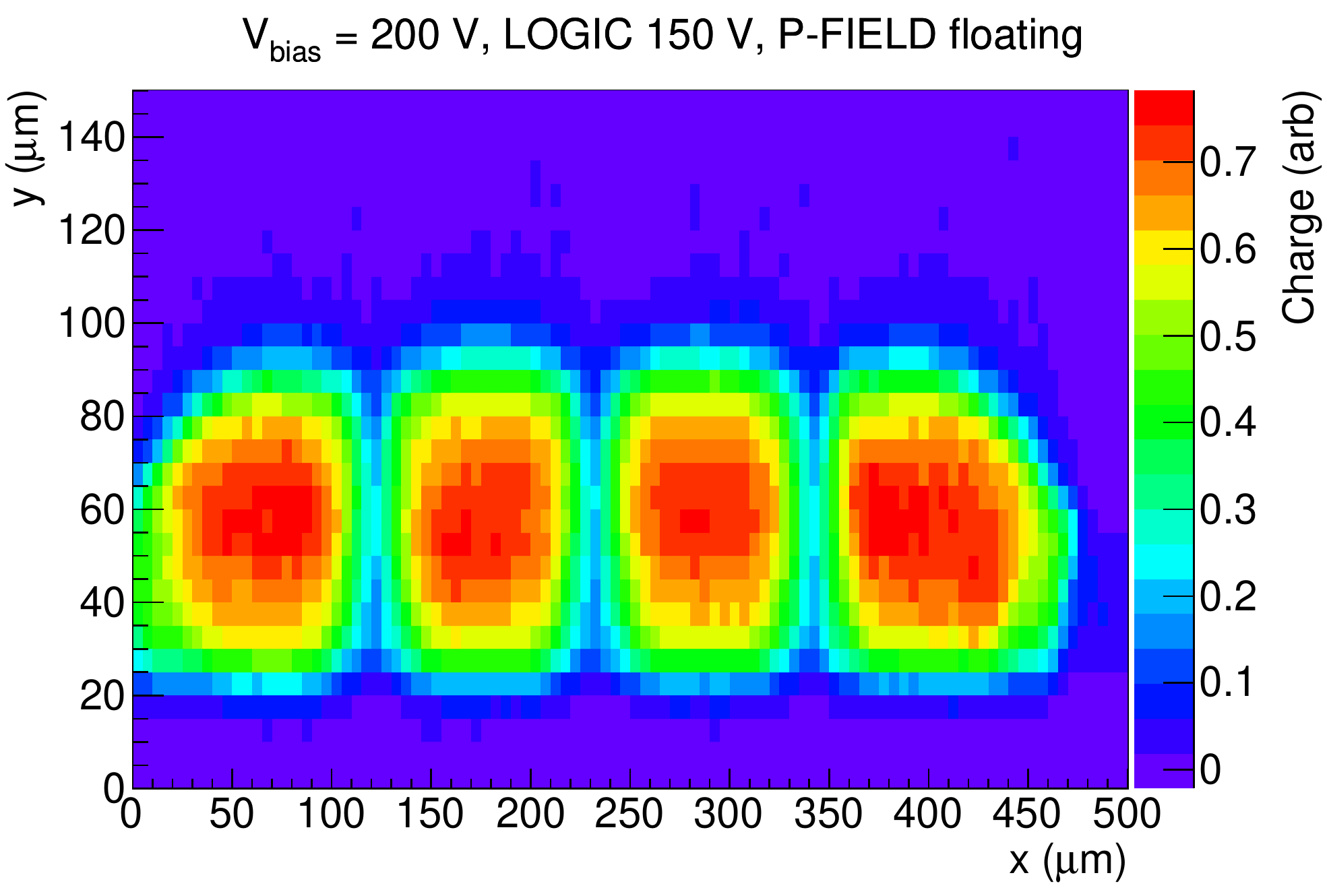}
\caption{$V_{\mathrm{bias}}=200\,\mathrm{V}$, $\mathrm{LOGIC}=150\,\mathrm{V}$}
\label{fig:gaps_150V}
\end{subfigure}%
\hfill
\begin{subfigure}[t]{0.47\textwidth}
\includegraphics[width=1\columnwidth,trim=0 0 0 0.15\columnwidth,clip]{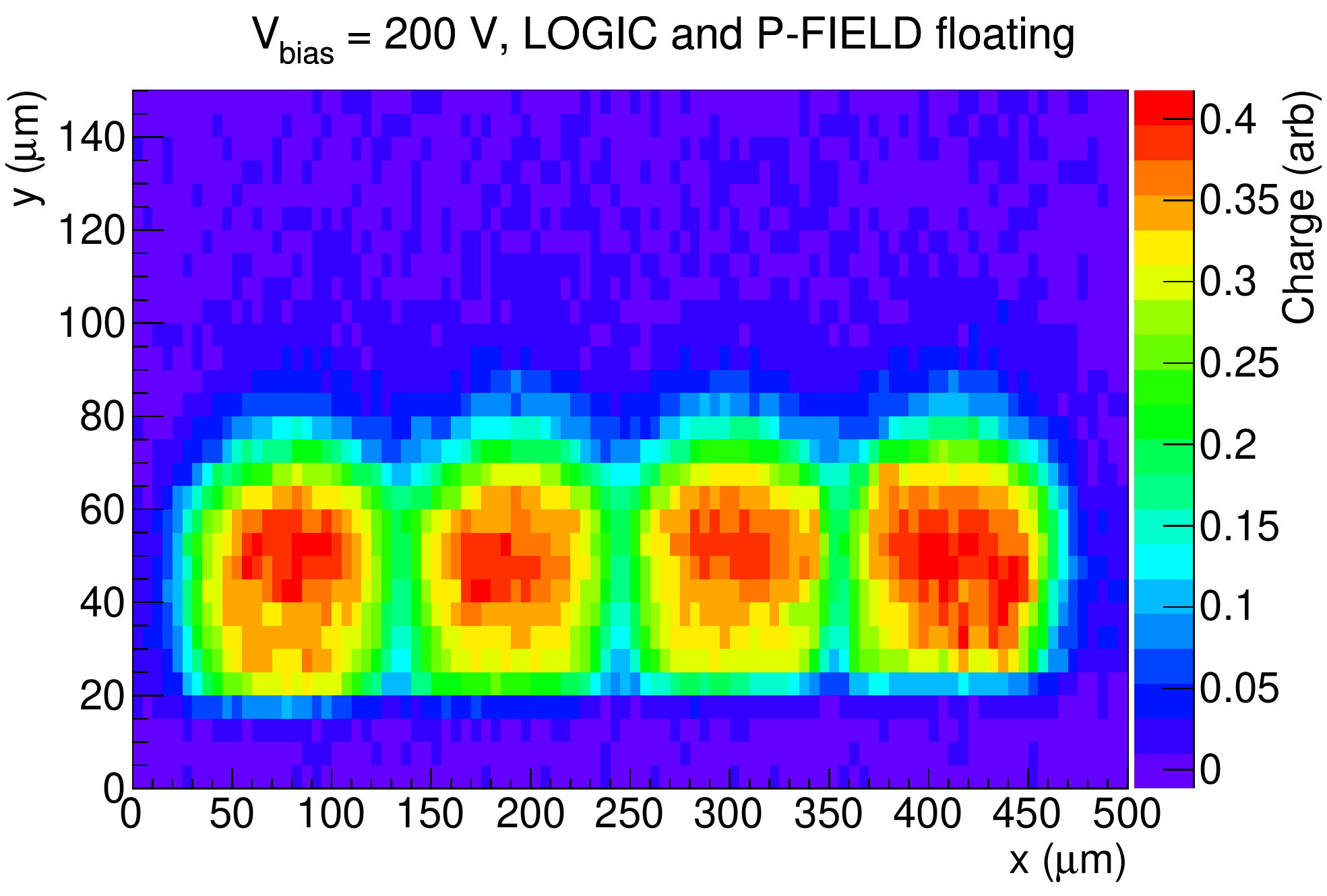}
\caption{$V_{\mathrm{bias}}=200\,\mathrm{V}$, LOGIC floating}
\label{fig:gaps_float}
\end{subfigure}%
\caption{Charge collection profiles in a $2\times 10^{14}\,\mathrm{n}_{\mathrm{eq}}/\mathrm{cm}^2$ irradiated sample for different LOGIC biasing configurations. The laser power was different between the measurements, hence the amount of collected charge differs.}
\label{fig:eff_gaps}%
\end{figure}

Before irradiation the efficiency gaps were not observed. This means that the properties of the BOX and the BOX-Si interface layer before irradiation enabled fast transport of charge carriers to the deep n-wells. Results of measurements shown in Figure \ref{fig:long_pulse} indicate that after irradiation charge carriers are temporarily trapped on the BOX-Si interface leading to slow charge collection. This may be the consequence of modified electric field due to charge trapped in the BOX layer and/or of the radiation induced defects on the BOX-Si interface slowing the charge transport towards the deep n-wells. Deeper understanding of this process is beyond of the scope of this work.

A further test of charge collection uniformity was carried out by biasing LOGIC to different bias voltages. In each measurement the deep n-well was always biased to $+200\,\mathrm{V}$ and the outer guard ring was grounded. LOGIC was then biased to either $+200\,\mathrm{V}$ (standard configuration), $+150\,\mathrm{V}$ or left floating. Two-dimensional charge collection profiles for different bias configurations were then recorded and are shown in Figure \ref{fig:eff_gaps}. All tests were done with a sample irradiated to a fluence of $2\times 10^{14}\,\mathrm{n}_{\mathrm{eq}}/\mathrm{cm}^2$.

The charge collection efficiency between the n-wells improves with respect to the standard configuration when LOGIC is biased to $+150\,\mathrm{V}$ or left floating.
In these cases the electric field lines bend more towards the deep n-wells and more electrons end their drift on the readout electrode, hence increasing the response uniformity. 
The efficiency gaps become less distinctive at higher neutron fluences.
The observed changes of the depleted depth for different biasing configurations in figure \ref{fig:eff_gaps} are not well understood, but may be related to screening of the electric field caused by positive charge accumulated in the BOX after irradiation.
A similar study was also carried out for the P-FIELD implant, however no large effects were observed in this case.

\section{Simulation of charge collection with KDetSim}
\label{sec:simulation}

Measured phenomena in the charge collection were qualitatively verified with the KDetSim simulation tool - a ROOT based library for simulation of charge transport in static detectors \cite{kdetsim}. 
For a given electrode configuration, space charge distribution and boundary conditions, KDetSim calculates the electric (weighting) field in the detector by solving the Poisson (Laplace) equation for the corresponding potential. Changes of the electric field due to injected charge carriers are neglected.
Injected charge is divided into buckets, which propagate through the detector as point charges. Moving buckets induce an electric current on the electrodes in accordance with the Shockley-Ramo theorem \cite{shockley, ramo}. Signals for electrons and holes are calculated separately and summed in the end. The induced charge is defined as an integral of the induced current from 0 to $25\,\mathrm{ns}$.  

\begin{figure}%
\centering
\includegraphics[width=0.5\columnwidth]{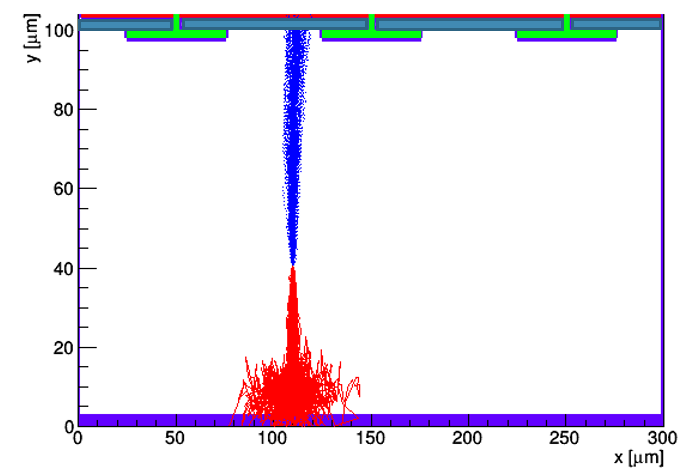}%
\caption{SOI device used in detector simulation. The detector details are described in the text. Shown are also drift paths of electrons (blue) and holes (red) after a point-like charge injection. Electron drift ends when they reach the BOX layer. Diffusion of holes in the undepleted part of the sensor can be observed in the lower part of the figure.}%
\label{fig:sim_layout}%
\end{figure}

In the simulation of the XTB02 chip several simplifications compared to the real device are assumed in order to reduce the complexity of the problem while still maintaining the main characteristics of the detector. 
The simulated device can be seen in Figure \ref{fig:sim_layout}. 
Detector calculation is done in two dimensions with three deep n-well implants ($50\,\upmu\mathrm{m}$ width, $100\,\upmu\mathrm{m}$ pitch) acting as readout electrodes. Above the deep n-wells a $1\,\upmu\mathrm{m}$ thick insulation layer and a $3\,\upmu\mathrm{m}$ thick silicon layer of BOX and LOGIC respectively are added. Charge carrier mobility in the BOX is assumed to be zero. A conductive back plane at a depth of $100\,\upmu\mathrm{m}$ serves as a back bias contact. Although the real device is $700\,\upmu\mathrm{m}$ thick and does not have a processed back plane this should not cause a significant difference, since the undepleted bulk is sufficiently conductive to be at the same electrical potential everywhere.

\begin{figure}%
\centering
\begin{subfigure}[t]{0.5\textwidth}
\centering
\includegraphics[width=0.9\columnwidth,trim=0 0 0 0.15\columnwidth,clip]{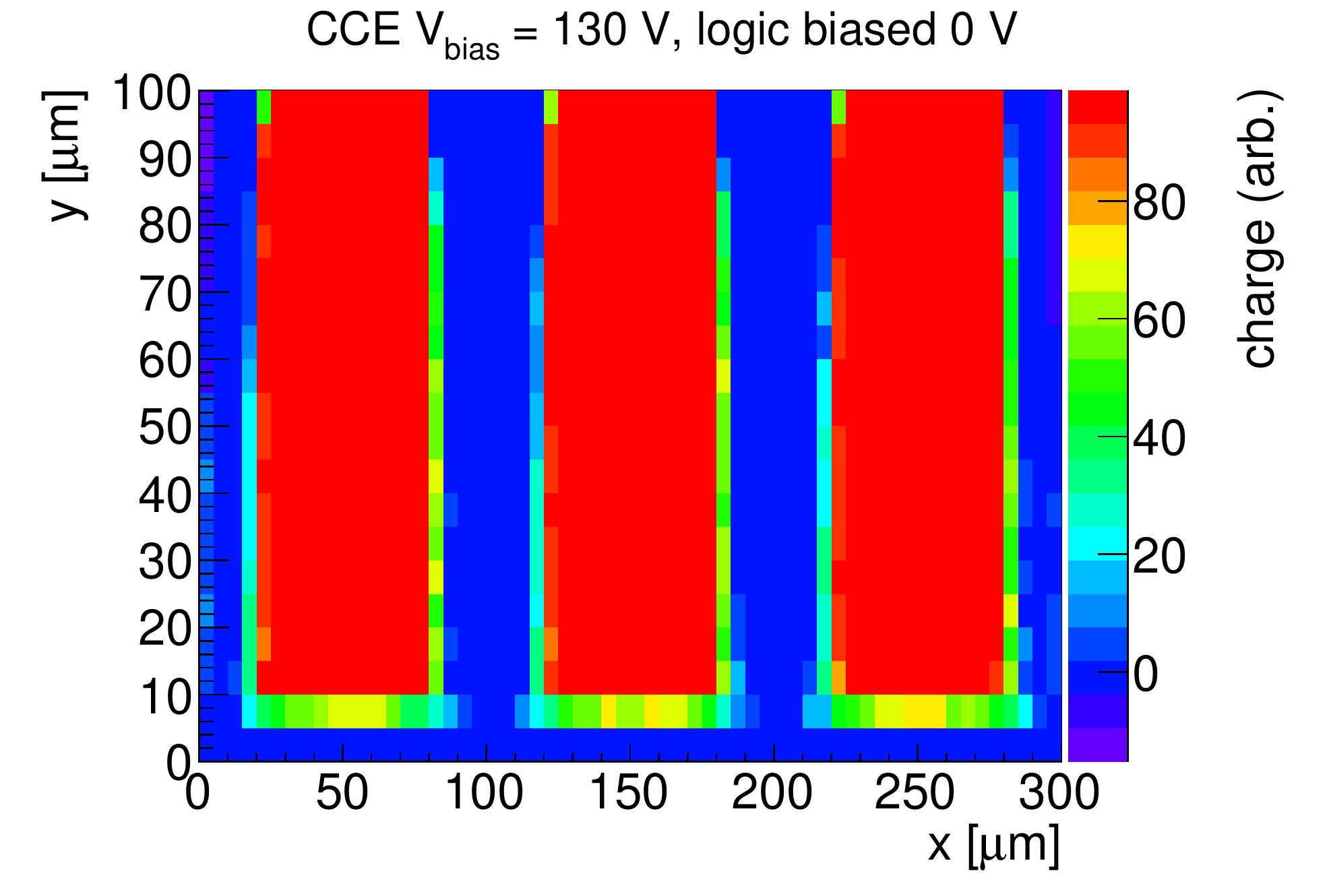}
\caption{$V_{\mathrm{back}}=-130\,\mathrm{V}$, $V_{\mathrm{implant}}=0\,\mathrm{V}$, $\mathrm{LOGIC}=0\,\mathrm{V}$}
\label{fig:sim_0V}
\end{subfigure}%
\\
\vspace{0.4cm}
\begin{subfigure}[t]{0.5\textwidth}
\centering
\includegraphics[width=0.9\columnwidth,trim=0 0 0 0.15\columnwidth,clip]{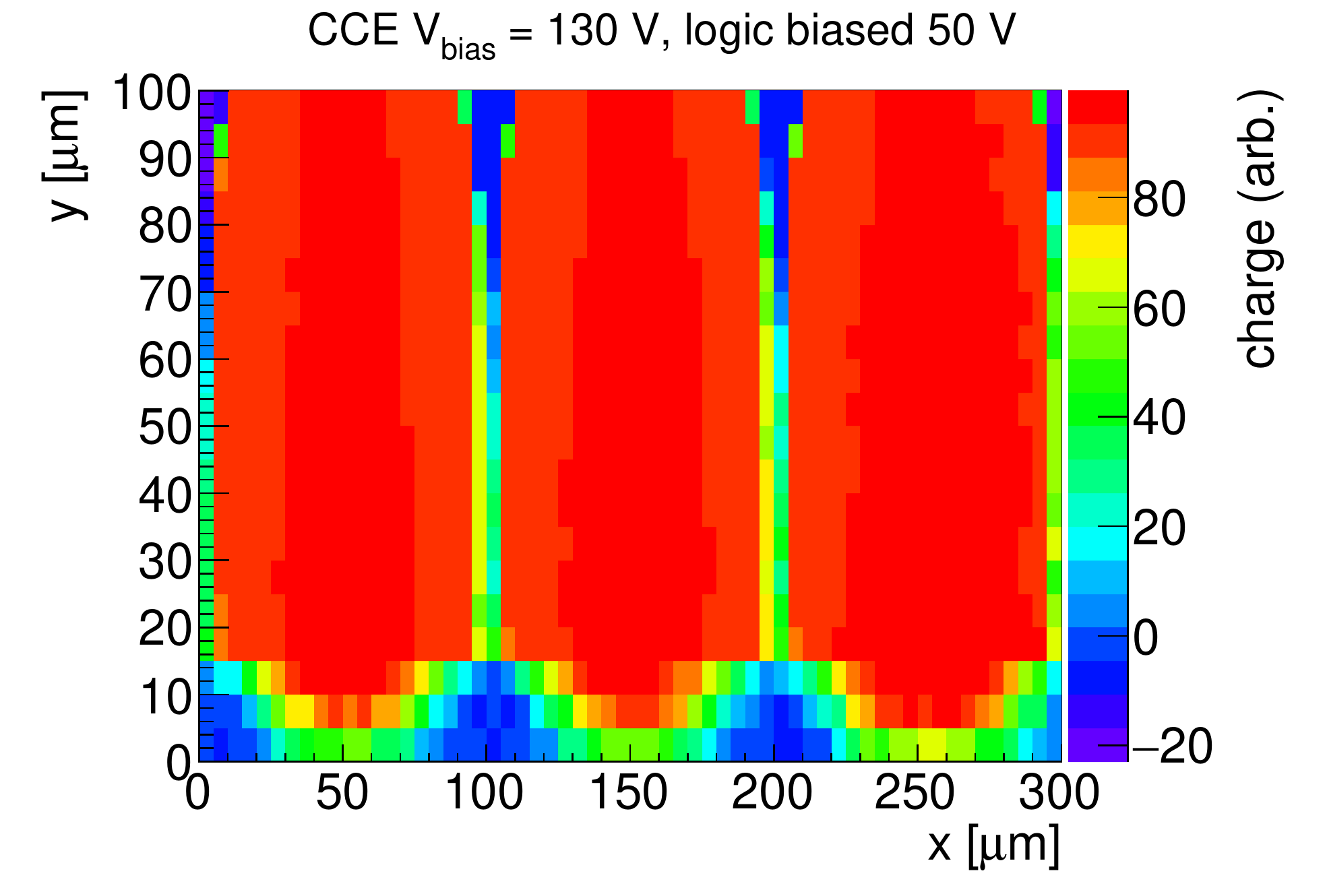}
\caption{$V_{\mathrm{back}}=-130\,\mathrm{V}$, $V_{\mathrm{implant}}=0\,\mathrm{V}$, $\mathrm{LOGIC}=-50\,\mathrm{V}$}
\label{fig:sim_50V}
\end{subfigure}%
\begin{subfigure}[t]{0.5\textwidth}
\centering
\includegraphics[width=0.9\columnwidth,trim=0 0 0 0.15\columnwidth,clip]{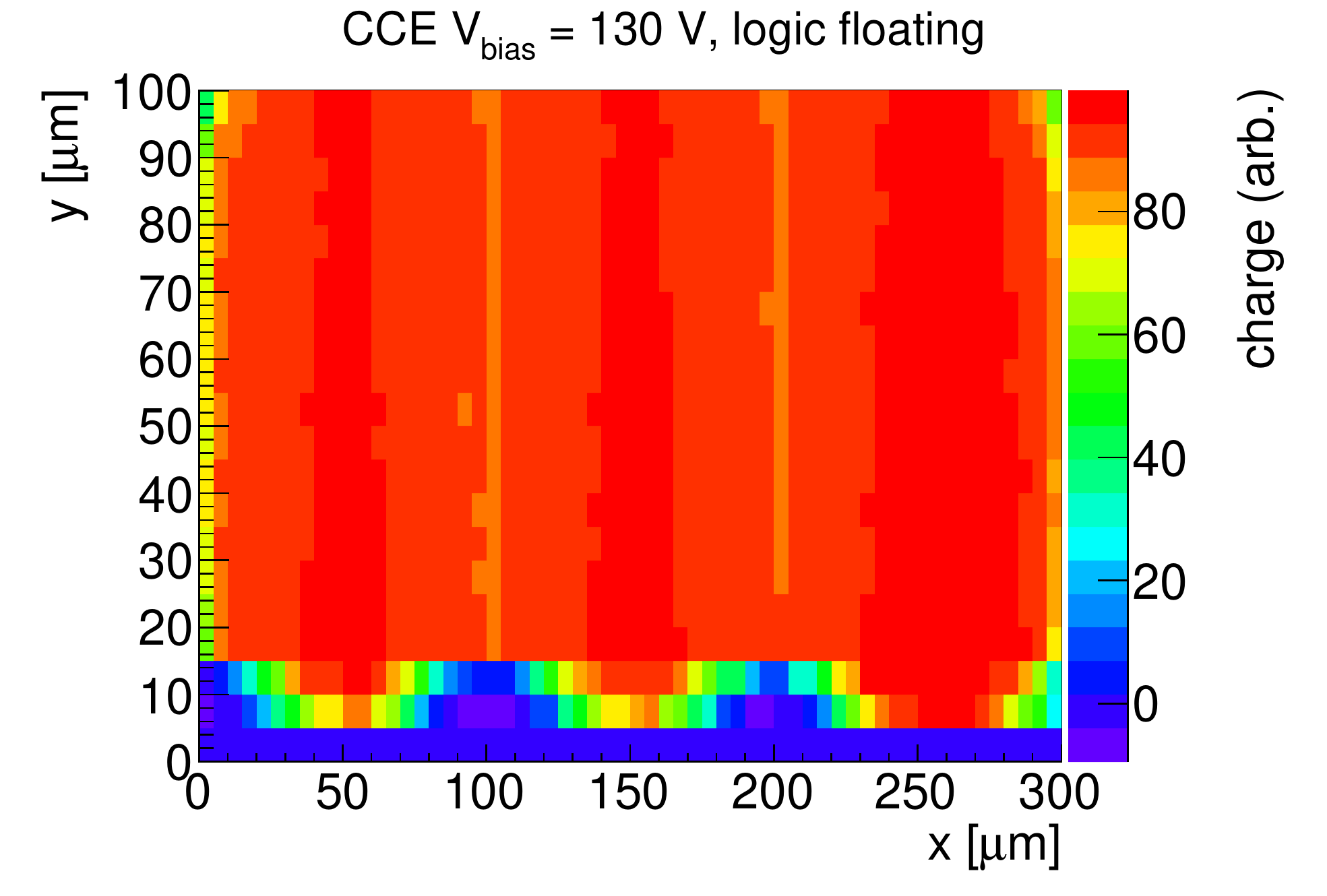}
\caption{$V_{\mathrm{back}}=130\,\mathrm{V}$, $V_{\mathrm{implant}}=0\,\mathrm{V}$, LOGIC floating}
\label{fig:sim_floating}
\end{subfigure}%
\caption{Simulated charge collection profiles for different biasing configurations of LOGIC. The results of the simulation are in a qualitative agreement with the measurements shown in Figure \ref{fig:eff_gaps}.}%
\label{fig:sim}%
\end{figure}

The simulation was performed with bias voltage between the grounded n-wells and the back plane set to $-130\,\mathrm{V}$, such that the sensor was not fully depleted.
Constant space charge concentration was assumed in the depleted region. Three electrodes with fixed electrical potentials were defined: the joined n-wells, the LOGIC and the back plane.
On the detector edges mirror boundary conditions with electric field lines parallel to the edges were set. Charge trapping on radiation induced defects was not taken into account, since it only becomes relevant on a time scale greater than the charge collection time in this simulation.
A two-dimensional charge collection scan was simulated for different bias voltages applied to the LOGIC electrode: $0\,\mathrm{V}$ (potential of the n-wells), $-50\,\mathrm{V}$ and LOGIC left floating. 
The step size in each directions was $5\,\upmu\mathrm{m}$. For each step a point-like injection of 100 buckets of charge was made. The resulting induced pulses were integrated over $25\,\mathrm{ns}$. The simulated charge collection profiles are shown in Figure \ref{fig:sim}. The results are qualitatively comparable to the E-TCT measurements. When LOGIC is biased to the same potential as the n-wells, efficiency gaps between the electrodes occur (Figure \ref{fig:sim_0V}). In configurations where electron drift towards the n-wells is more favored, the efficiency gaps between the electrodes are reduced (LOGIC at $-50\,\mathrm{V}$, Figure \ref{fig:sim_50V}) or disappear completely (LOGIC floating, Figure \ref{fig:sim_floating}).
Note that in this simulation the drift of charge is stopped when it reaches the Si-BOX interface. The properties of the interface enabling the transport of charge to the n-wells are not considered so the efficiency gaps are seen also in what would represent a model of unirradiated detector.

\section{Conclusions}

The paper reports about an investigation of charge collection properties before and after neutron irradiation in a CMOS pixel detector prototype produced on a $100\,\Omega\cdot\mathrm{cm}$ p-type substrate in SOI technology by XFAB. The depleted depth was estimated by E-TCT. 
At $300\,\mathrm{V}$ bias voltage the thickness of charge collection layer initially increased with irradiation from $50\,\upmu\mathrm{m}$ before irradiation to $160\,\upmu\mathrm{m}$ after a neutron fluence of $5\times 10^{14}\,\mathrm{n}_{\mathrm{eq}}/\mathrm{cm}^2$.  
At higher fluences the depleted depth falls but even at highest fluence of $1\times 10^{16}\,\mathrm{n}_{\mathrm{eq}}/\mathrm{cm}^2$ it remains larger than $30\,\upmu\mathrm{m}$.
These changes are a consequence of radiation induced removal of initial dopants and introduction of stable deep acceptors.
The parameters describing the evolution of $N_{\mathrm{eff}}$ with fluence were extracted from the fit of measured data. The value of acceptor removal constant was larger than the constant estimated with similar method on HV-CMOS devices made on lower resistivity substrate \cite{ChessRemoval}. This is in agreement with the hypothesis that the removal constant is larger in material with a higher initial resistivity.   
A study of charge collection uniformity within the pixels revealed efficiency gaps on pixel edges. E-TCT tests with different detector biasing configurations, as well as computer simulations, showed that they appear due to a parasitic AC coupled charge collection by the area above the BOX layer not covered with the deep n-well. It was shown that with an appropriate biasing scheme and/or larger n-well fill factor this effect might be greatly reduced.

\acknowledgments

The authors would like to thank the crew at the TRIGA reactor in Ljubljana for help with irradiations of detectors.
The authors acknowledge the financial support from the Slovenian Research Agency (research core funding No. P1-0135 and project ID PR-06802).

\end{document}